\pdfoutput=1
\PassOptionsToPackage{table}{xcolor}
\documentclass[sigconf,preprint,natbib=true]{acmart}

\usepackage{amsthm}
\usepackage{xspace}
\usepackage{amsmath}
\usepackage{pgfplots}
\usepackage{enumitem}
\usepackage{adjustbox}
\usepackage{subcaption}
\usepackage{booktabs, multirow}
\usepackage{mathtools}
\usepackage{comment}
\usepackage{csquotes}
\usepackage{pifont}
\usepackage{tikz}
\usepackage{soul}
\usepackage{breqn}
\usepackage[ruled,vlined]{algorithm2e}
\usepackage[flushleft, para]{threeparttable}

\newcommand*{\TallestContent}{APLT}

\newcommand*{\ra}[1]{\overrightarrow{\makebox[2.0em]{$#1$\vphantom{\TallestContent}}}}
\newcommand*{\la}[1]{\underleftarrow{\makebox[2.0em]{$#1$\vphantom{\TallestContent}}}}
\newcommand*{\ga}[1]{\overleftarrow{\makebox[2.0em]{$#1$\vphantom{\TallestContent}}}}

\definecolor{ylgnbu1}{RGB}{255, 204, 204}
\definecolor{ylgnbu2}{RGB}{255, 102, 102}
\definecolor{ylgnbu3}{RGB}{255, 51, 51}
\definecolor{ylgnbu4}{RGB}{153, 0, 0}
\definecolor{ylgnbu5}{RGB}{51, 0, 0}
\definecolor{ylgnbu6}{RGB}{0, 153, 0}

\newenvironment{customlegend}[1][]{%
\begingroup
\csname pgfplots@init@cleared@structures\endcsname
\pgfplotsset{#1}%
}{%
\csname pgfplots@createlegend\endcsname
\endgroup
}%
\def\addlegendimage{\csname pgfplots@addlegendimage\endcsname}

\setcopyright{acmcopyright}
\copyrightyear{}
\acmYear{}
\acmDOI{}

\acmConference[xxx]{xxx}{xxx}{xxx}
\acmPrice{15.00}
\acmISBN{978-1-4503-XXXX-X/18/06}

\acmBooktitle{}

\newcommand{\pdu}{\textsf{PDU}\xspace}
\newcommand{\easer}{EASE$^R$\xspace}

\begin{document}

\title{Post-hoc Selection of Pareto-Optimal Solutions in Search and Recommendation}

\author{Vincenzo Paparella}
\email{vincenzo.paparella@poliba.it}
\affiliation{%
  \institution{Polytechnic University of Bari}
  \city{Bari}
  \country{Italy}
}

\author{Vito Walter Anelli}
\email{vitowalter.anelli@poliba.it}
\affiliation{%
  \institution{Polytechnic University of Bari}
  \city{Bari}
  \country{Italy}
}

\author{Franco Maria Nardini}
\email{francomaria.nardini@isti.cnr.it}
\affiliation{%
  \institution{ISTI---CNR}
  \city{Pisa}
  \country{Italy}
}

\author{Raffaele Perego}
\email{raffaele.perego@isti.cnr.it}
\affiliation{%
  \institution{ISTI---CNR}
  \city{Pisa}
  \country{Italy}
}

\author{Tommaso Di Noia}
\email{tommaso.dinoia@poliba.it}
\affiliation{%
  \institution{Polytechnic University of Bari}
  \city{Bari}
  \country{Italy}
}

\settopmatter{printacmref=false}
\renewcommand\footnotetextcopyrightpermission[1]{}

\newcommand{\msn}{\texttt{MSN30K}\xspace}


\begin{abstract}
Information Retrieval (IR) and Recommender Systems (RS) tasks are moving from computing a ranking of final results based on a single metric to multi-objective problems. Solving these problems leads to a set of Pareto-optimal solutions, known as Pareto frontier,  in which no objective can be further improved without hurting the others.
In principle, all the points on the Pareto frontier are potential candidates to represent the best model selected with respect to the combination of two, or more, metrics.
To our knowledge, there are no well-recognized strategies to decide which point should be selected on the frontier. 
In this paper, we 
propose a novel, post-hoc, theoretically-justified technique, named ``Population Distance from Utopia'' (\pdu),  to identify and select the one-best Pareto-optimal solution from the frontier. In detail, \pdu analyzes the distribution of the points by investigating how far each point is from its utopia point (the ideal performance for the objectives). The possibility of considering fine-grained utopia points allows \pdu to select solutions tailored to individual user preferences, a novel feature we call ``calibration''.
We compare \pdu against existing state-of-the-art strategies through extensive experiments on tasks from both IR and RS.
Experimental results show that \pdu and combined with calibration notably impact the solution selection.
Furthermore, the results show that the proposed framework selects a solution in a principled way, irrespective of its position on the frontier, thus overcoming the limits of other strategies.
\end{abstract}


\maketitle

\vspace{-0.3em}

\section{Introduction}
\label{sec:intro}
Many tasks in Information Retrieval (IR)  and Recommender Systems (RS) involve the optimization of multiple objective functions. 
As an example, consider the IR task of \emph{diversifying search results} where, given a user query, we require the IR system to return a list of results that are both \emph{relevant} for the user and \emph{diverse} concerning the possible ``facets'' of the query \cite{INR-040}. Addressing this task asks for designing a two-objective  ranking function comprehensively maximizing both the relevance and the diversity of the result list. 
The same considerations can be made in RS. Despite the accuracy of recommendation being considered the gold measure to assess the quality of suggestions, over the last years, RSs have been required to meet other \textit{beyond-accuracy} metrics to avoid obvious~\cite{DBLP:conf/recsys/VargasC11} and unfair~\cite{wu2022multifr} recommendations. Therefore, the choice of a recommendation model and its setting entail several criteria 
leading to a trade-off among them, resulting in a non-trivial challenge.

Multi-objective Optimization (MOO) recently attracted several interesting IR and RS contributions~\cite{wu2022multifr, DBLP:conf/wsdm/GeZYPHHZ22, DBLP:conf/wsdm/StamenkovicKAXK22}. MOO deals with \emph{Pareto optimality}, i.e., the identification of  solutions  
where no objective can be further improved without damaging the others. 
Pareto-optimal solutions are in turn collected in the so-called \emph{Pareto Frontier}, a set of (possibly infinite) non-dominated solutions.

Existing approaches for MOO can be classified into two categories: i) \emph{heuristic search}
and, ii) \emph{scalarization}. In the first category, multi-objective evolutionary algorithms are used to ensure that the emerging solutions are not dominated by
each other, even if they can still be dominated by Pareto-optimal solutions not visited by the algorithm~\cite{DBLP:conf/dagstuhl/2008moo, DBLP:conf/recsys/RibeiroLVZ12}. 
In the second category, scalarization methods aggregate multiple objectives into one objective, possibly guaranteeing Pareto optimality. 
Scalarization approaches can exploit \emph{model aggregation} techniques combining the output of different models trained on the single objectives. Alternatively, \emph{label aggregation} techniques combine the labels of the training samples a priori, and the optimization is performed using the aggregated label. 
Aggregation techniques may involve the setting of the importance or priority of the different objectives by weighting each objective through a scalar function, e.g., Linear Scalarization~\cite{moosurvey}, Weighted Chebyshev~\cite{1084969}. 
Conversely, some techniques work by constraining the objectives of the problem, e.g., $\epsilon$--Constraint~\cite{4308298} leading to a unique non-dominated solution.

Pareto optimality is commonly achieved by many different Pareto-optimal solutions. However, IR and RS MOO tasks generally require identifying a single Pareto-optimal solution to be deployed in the system. To the best of our knowledge,  no strategies  specifically tailored to   IR and RS tasks have been previously proposed~\cite{wu2022multifr}. The state-of-the-art techniques from MOO theory are in fact aimed at identifying a set of Pareto-optimal solutions, without addressing the problem of \emph{post-hoc} choosing one among the---possibly many---solutions identified for the IR and RS tasks. Indeed, many works in the IR and RS literature, although exploiting the techniques discussed above, 
do not either: i) consider the problem of selecting a single best solution to the multi-objective problem or, ii), 
discuss the criteria adopted to select a single Pareto-optimal solution~\cite{DBLP:journals/ijon/ZhengW22}.

In this paper, we fill this gap by introducing ``Population Distance from Utopia'' (\pdu), a novel post-hoc flexible strategy for selecting \textbf{one}---\textbf{best}---Pareto-optimal solution among the ones lying in the Pareto frontier for IR and RS tasks. \pdu relies on the observation that the Pareto-optimal point coordinates are an aggregation---usually the mean---of the model performance for each sample, i.e., queries in IR and users in RS, on multiple objectives. \pdu exploits the notion of ``Utopia point'' as the ideal optimization target. Differently from the methods from MOO theory, which are devised to solely consider the mean performance values when selecting a single Pareto-optimal solution, \pdu~is designed to set a utopia point for each sample of the dataset. This feature allows choosing a solution not only based on the ``global''  performance achieved by the IR/RS model, but also in a more fine-grained resolution that now considers multiple quality criteria that~are expressed on a sample level. We call this feature ``calibrated'' selection. In detail, the novel contributions of this paper are:
\begin{itemize}[leftmargin=*]
    \item We formally introduce \pdu as a novel technique that allows one to select, in a principled way,  the best Pareto-optimal solution previously identified by a state-of-the-art MOO technique.
    \item We provide a thorough comparison of \pdu against state-of-the-art selection strategies. The analysis shows  that \pdu is the only selection method that allows identifying a ``calibrated'' solution, i.e., based on ideal targets expressed on a sample level.
    \item We experimentally compare \pdu against~state-of-the-art strategies on well-known IR and RS tasks by exploiting public data. The results show that, unlike  other methods,  \pdu  can identify Pareto-optimal solutions regardless of their position on the frontier. Moreover, \pdu  calibration can lead to the selection of significantly different trade-offs.
    \item We release a GitHub repository\footnote{\scriptsize{\url{https://anonymous.4open.science/r/Selection-Pareto-Optimal-Solutions-IR-RS-56A4/}}} for our implementation of \pdu and the state-of-the-art competitors as well as the data used in the experiments to allow a full reproducibility of the results. 
\end{itemize}



\section{Multi-Objective Optimization}
\label{sec:background}
A Multi-Objective Optimization Problem (MOOP) \cite{moosurvey} is defined~as:
\begin{equation}
\label{eq:p1}
\begin{aligned}
\min_{\mathbf{x}} \quad & \text{\textit{\textbf{f}}}(\mathbf{x})=\left\{f_{1}(\mathbf{x}), f_{2}(\mathbf{x}), \ldots, f_{k}(\mathbf{x})\right\} \\
\textrm{subject to} \quad & \mathbf{x} \in \mathcal{X}.
\end{aligned}
\end{equation}

The vector $\mathbf{x} \in \mathbb{R}^n$ is formed by $n$ independent variables called \textit{decison variables}. The set $\mathcal{X} \subseteq \mathbb{R}^n$, generally known as \textit{feasible set}, is defined by a set of equality and inequality constraints such as $\{\mathbf{x}\,|\,g_j(\mathbf{x}) \leq 0, j=1,2,\dots, l; \, \vee \, h_i(\mathbf{x})=0, i = 1,2,\dots, e\}$. The vector of functions \textit{\textbf{f($\cdot$)}} is composed by $k \geq 2$ scalar \textit{objective functions}  $f_i: \mathcal{X} \rightarrow \mathbb{R} \, \textrm{with} \,  i=1,\dots, k$. In multi-objective optimization, the space $\mathbb{R}^k$ is known as \textit{objective function space}.  

Note that a MOOP as defined in Equation~(\ref{eq:p1}) supposes that improving one objective means minimizing it. However, an objective $f(\cdot)$ of the vector of functions \textit{\textbf{f($\cdot$)}} might have either a maximization or a minimization goal. In this sense, maximization of the function $f(\cdot)$ may be readily rewritten as $-f(\cdot)$ to meet Equation~(\ref{eq:p1}).

\vspace{1mm}
\noindent \textbf{Pareto Optimality.}
In a single-objective optimization problem, the optimal solution is defined as the objective function value that proves the scalar relation ``less than or equal'' ($\leq$). In contrast, in a MOOP, since typically there is no single global solution, it is impossible to determine a set of points that all fit a predetermined definition for an optimum. Hence, it is usually adopted the concept of \textit{Pareto optimality} which leverages on the \textit{Pareto dominance} relation~\cite{DBLP:journals/coap/LuckenBB14}. A vector $\mathbf{x}^\star$ Pareto-dominates vector $\mathbf{x}$, denoted by $\mathbf{x}^\star \prec \mathbf{x}$, if and only if
    $\exists j \in \{1, \dots, k\}  \, | \,\, f_j(\mathbf{x^\star}) < f_j(\mathbf{x})  \,
    \textrm{and} \,\,
    f_i(\mathbf{x^\star}) \leq f_i(\mathbf{x}) \,\, \forall i \in \{1, \dots,j-1,j+1,\ldots, k\}.$
We also write that, a solution $\mathbf{x^\star} \in \mathcal{X}$ is Pareto optimal if there does not exist another solution $\mathbf{x} \in \mathcal{X}$ such that $\mathbf{f}(\mathbf{x}) \prec \mathbf{f}(\mathbf{x^\star})$.
In other words, a point is Pareto optimal if there is no other point that improves at least one objective function without hurting another one. Then, solving the problem in Equation \eqref{eq:p1} means finding the solutions $\mathbf{x} \in \mathcal{X}$ such that their images $\mathbf{f}(\mathbf{x})$ are not Pareto-dominated by any other vector in the feasible set. The set of non-Pareto-dominated solutions $P^\star~\subseteq~\mathcal{X}$ is called Pareto-optimal set in the feasible set, that is formally defined as $P^\star := \{ \mathbf{x^\star} \in \mathcal{X} | \neg \exists\, \mathbf{x}\in\mathcal{X}\,\, \textrm{s.t.}\,\, \mathbf{x} \prec \mathbf{x^\star} \}$.
The image of the Pareto-optimal set $P^\star$ in the objective function space is called the Pareto frontier, i.e., $PF^\star := \{ \text{\textit{\textbf{f}}}(\mathbf{x^\star})\,|\, \mathbf{x^\star} \in P^\star \}$.

\vspace{1mm}
\noindent \textbf{Utopia and Nadir Points.}
Once a solution $P^\star$ for the problem in Equation \eqref{eq:p1} is obtained, the decision-making process requires the selection of a single optimal solution from the Pareto frontier. Generally, the \textit{utopia point} helps to implement this process~\cite{moosurvey}. A point $\textit{\textbf{f}}^{\,\diamond} \in \mathbb{R}^k$ is a utopia point if and only if $\textit{f}^{\,\diamond}_i =
\min_{\mathbf{x}} \,
{\textit{f}_i (\mathbf{x})\,|\,\mathbf{x} \in \mathcal{X}}\,\, \forall i \in \{1, 2, \dots, k\}$.
Generally, the utopia point is the \textit{ideal} point in $\mathbb{R}^k$ that is unattainable. Hence, a common approach consists in reaching the \textit{closest} solution to the utopia point as the best one, where, in most of the cases, 
the term \textit{closest} refers to the solution which minimizes the Euclidean distance to the utopia point. However, it is not necessary to restrict closeness to the case of a Euclidean norm~\cite{moosurvey}.

Along with the utopia point, the \textit{nadir point} also helps select a solution from the Pareto frontier. Dually to the utopia point, the nadir point represents the point in the objective function space having the worst possible values for each objective.
A point $\textit{\textbf{f}}^{\,\triangle} \in \mathbb{R}^k$ is a nadir point if and only if $\textit{f}^{\,\triangle}_i =
\max_{\mathbf{x}} \,
{\textit{f}_i (\mathbf{x})\,|\,\mathbf{x} \in \mathcal{X}}\,\, \forall i \in \{1, 2, \dots, k\}$.
Compared to the utopia point, determining the nadir point can be challenging, even for simple problems~\cite{DBLP:journals/csur/LiY19}. 
\section{Background}
\subsection{Selection Strategies}
\label{sec:baseline}
The Pareto frontier consists of a set of equally optimal solutions. 
Some methods to select a single Pareto-optimal solution assume the existence of a decision maker~\cite{DBLP:journals/jors/Leake01}. These methods are known as \textit{Multi-Criteria Decision Making} (MCDM) strategies, where a decision-maker has knowledge of the preferences (hierarchy) among the objectives. However, decision-makers do not always know how to weigh the different objectives~\cite{DBLP:conf/ppsn/BrankeDDO04}. Moreover, in some cases, the complexity of the problem  makes it difficult for a human decision-maker to evaluate and compare different options comprehensively. Conversely, mathematical methods can provide consistent, objective, and impartial decision-making approaches. In this work, we focus and outline mathematical strategies for selecting a solution from the Pareto frontier, i.e., strategies applicable in the absence of  \enquote{a priori knowledge} that can feed an MCDM strategy.
\subsubsection{Knee Point}\label{sec:knee_point}
The \textit{Knee Point}~\cite{DBLP:conf/ppsn/BrankeDDO04} strategy aims to identify a knee of the Pareto frontier. 
The rationale is that solutions different from the knee point would exhibit limited improvement for one objective and a substantial deterioration for the others. As described by~\citet{DBLP:conf/ppsn/BrankeDDO04}, 
these strategies were born as a variation of multi-objective evolutionary algorithms to find the knee regions on the Pareto frontier. Consequently, when other algorithms compute the Pareto Frontier, the extracted knee region may not have a knee-featured shape, thus making this strategy less convenient. Several methods to identify the knee point are proposed in the literature, mainly differing for the number of objectives.

\vspace{1mm}
\noindent
\textbf{Angle-based method (A-KP).}
When dealing with two objectives, the reflex angle between the slopes of the two vectors through a point $B=(x_i,y_i)$ and its two neighbors, i.e., $A=(x_{i-1},y_{i-1})$ and $C=(x_{i+1},y_{i+1})$, on the Pareto Frontier can be considered as an efficient indication of whether the point can be classified as a knee~\cite{DBLP:conf/ppsn/BrankeDDO04}.
\textit{The Pareto-optimal point having the maximum reflex angle computed from its neighbors is considered the knee}~\cite{doi:10.1080/0305215X.2010.548863}. If no neighbor to the left (right) is found, a vertical (horizontal) line is used to
calculate the angle.
Even though this method is efficient in a two-dimensional scenario, it becomes impractical for more than two objectives, especially for the choice of neighbors.

\vspace{1mm}
\noindent
\textbf{Utility-based method (U-KP).}
A valid alternative to overcome the limitation of the angle-based method is adopting a marginal utility function. 
Let us consider a set of $n$ objective functions $f(\cdot)$ and $m$ sets of $n$ uniformly distributed weights $\mathbf{w}$, with $w_i \in [0,1]$ such that $\sum_i w_i = 1$~\cite{DBLP:conf/ppsn/BrankeDDO04}.
The resulting utility function is then $U(\mathbf{x},\mathbf{w})=\sum_i w_i\cdot f_i(x)$.
The solution having the minimum utility value (Pareto-optimal solution) for most weight configurations is the knee point.
\subsubsection{Hypervolume}\label{sec:hypervolume}
The \textit{Hypervolume}~\cite{DBLP:conf/emo/ZitzlerBT06} strategy was first introduced to compare the quality of different Pareto frontiers~\cite{DBLP:conf/emo/Fleischer03}. However, by computing the hypervolume of each solution on the Pareto frontier, this strategy can be straightforwardly exploited to select the best solution from the set~\cite{DBLP:journals/ijon/ZhengW22}. 
Given a Pareto-optimal solution $\mathbf{x^\star} \in \mathbb{R}^k$, 
a reference point $\mathbf{r}\in\mathbb{R}^k$, and the Lebesgue measure $\lambda$, the hypervolume $\mathcal{HV}$ of $\mathbf{x^\star}$ with respect to $\mathbf{r}$ is:
\begin{equation}\label{eq:hv}
\mathcal{HV} = \lambda(\{\mathbf{x}\in\mathbb{R}^k\,|\,\mathbf{x^\star}\prec \mathbf{x}\prec\mathbf{r}\}).
\end{equation}
The $\mathcal{HV}$ value is the volume of the hypercube determined by the solution $\mathbf{x^\star}$ and the reference point $\mathbf{r}$. 
\textit{The Pareto-optimal point having the maximum hypervolume is the selected one}. 


\subsubsection{Other Techniques}\label{sec:other_techniques}
Other simpler techniques that have been used for selecting a solution from the Pareto frontier are the \textit{Euclidean Distance} and the \textit{Weighted Mean}~\cite{DBLP:journals/isci/NoiaRTS17, DBLP:journals/kbs/WangGLY16}.
The Euclidean Distance (\textit{ED}) is computed between each solution on the Pareto frontier and the utopia point:
\begin{equation}
    ED(\mathbf{x^\star}) = |\textit{\textbf{f}} (\mathbf{x^\star}) - \textit{\textbf{f}}^{\,\diamond}| = \left( \sum_{i=1}^k (f_i(\mathbf{x^\star})- f_i^{\,\diamond})^2\right)^{\frac{1}{2}}.
\end{equation}
\textit{The Pareto-optimal point having the minimum Euclidean distance is the selected solution.}
Instead, the Weighted Mean (\textit{WM}) requires setting the importance of each objective through a set of weights. \textit{Among all the Pareto-optimal points, the point maximizing the weighted mean corresponds to the selected solution.}

\subsection{Related Work on MOO for IR and RS}
\label{sec:related}
While a discussion of the state-of-the-art selection approaches has been provided in Section \ref{sec:baseline}, we now briefly summarize the main contributions of MOO in IR and RS. Previous works investigate the introduction of multiple criteria in IR systems, e.g., in web search and recommendation \cite{dai2011learning,10.1145/2124295.2124350,10.1145/2187836.2187894,10.1145/1963405.1963459, 10.1145/2911451.2914708,10.1145/2348283.2348385}, and product search \cite{10.1145/3077136.3080838,10.1145/2396761.2398671}. \citet{DBLP:conf/www/CarmelHLL20} propose Stochastic Label Aggregation (SLA), a technique that perform label aggregation by randomly selecting a label per training example according to a given distribution over the labels. 
In RS, \citet{DBLP:conf/recsys/LinCPSXSZOJ19} propose a scalarization based Pareto-Efficient Learning-To-Rank (PE-LTR) framework 
by deriving the conditions for the weighted sum weights that ensure the solution to be Pareto efficient.
In the RS area, MOO techniques are routinely exploited for optimizing multiple fairness criteria beyond relevance. \citet{DBLP:conf/wsdm/GeZYPHHZ22} propose a fairness-aware RS based on multi-objective reinforcement learning, simultaneously optimizing clickthrough rate (CTR), as a signal for relevance, and item exposure, as a signal for fairness. Moreover, \citet{wu2022multifr} employ scalarization to optimize accuracy along with both provider and consumer fairness. \citet{DBLP:conf/sigir/NaghiaeiRD22} also integrate fairness constraints from a consumer and producer-side into a 
re-ranking approach. 



\section{Population Distance from Utopia}
\label{sec:pdu}
Driven by the goal of overcoming the limitations of the other methods in a principled way for IR and RS, we propose \pdu (Population Distance from Utopia), a selection strategy taking into account the distance of the query/user metric from the utopia point.

Our intuition starts from the observation that in a search and/or recommendation scenario, the Pareto frontier is populated by points representing aggregated results (usually, they represent the average value) on metrics referring to a set of experiments. For instance, in a RSs setting, we could have a frontier representing the values of two metrics: \textit{nDCG}, to measure the accuracy of the model, and \textit{Intralist Diversity} (\textit{ID}), to measure the diversity in the list of recommended items. Each point on the frontier may represent the corresponding values of \textit{nDCG} and \textit{ID} for a specific configuration of the hyperparameters. It is worth noticing that these values are computed as the value of the given metric averaged on all the system users. Suppose we focus instead on the point representing the single user. In that case,  we may also reconsider the notion of utopia point in this more fine-grained view and adapt it to generalize with respect to the single user. The same observations hold in a search setting where we have queries instead of users. The questions leading our proposal are then: i)\textit{What happens if we focus our analysis on the original points instead of their aggregated  representation?} ii) \textit{Can we characterize each of these fine-grained points and exploit a generalized definition of  utopia point that considers even the single user/query?}\\
We start by defining a generalized version of the utopia point. 

A point $\textit{\textbf{f}}^{\,\circ}$ in the \textit{objective function space} $\mathbb{R}^k$ is a \textit{\textbf{generalized utopia point}} if and only if $\textit{f}^{\,\circ}_i ={\textit{h}_i (\mathbf{x})\,|\,\mathbf{x} \in \mathcal{X}}\,\, \forall i \in \{1, 2, \dots, k\}$.
In our definition, $h_i$ is a function that considers the characteristics of the original data and returns a desired but unattainable utopia value for the $i$-th metric. For a (non-generalized) utopia point $f^\diamond$, we have $\textit{h}_i = \min_{\mathbf{x}} \, \textit{f}_i (\mathbf{x})$. Its definition can be driven both by system or dataset properties and by the choices of the system designer. For instance, in Section~\ref{sec:expsc}, we define $h_2$ (see Equation \eqref{eq:apltutopia}) to quantify the users' popularity tendencies stemming from their past interactions with the items in a recommendation scenario.
Given a Pareto-optimal solution $\mathbf{x^\star} \in \mathbb{R}^k$, we can assume that it is the image of an aggregation function applied to a set of $m$ points $\mathbf{x}_j$ in $\mathbb{R}^k$, with $j \in \{1,\dots, m\}$. In our previous example, the points represent the values of the pairs $\langle nDCG, ID\rangle$ (with $k=2$) for the $m$ users in the system. Suppose a generalized utopia point $\textit{\textbf{f}}^{\,\circ}_j \in \mathbb{R}^k$, with $j \in \{1,\dots, m\}$, is associated to each point $\mathbf{x}_j$.
\begin{definition}
The \textit{Population Distance from Utopia} (\pdu) is:
\begin{equation}\label{eq:pdu}
    \pdu = \log \left( \sum_{j=1}^m e(\textit{\textbf{f}}^{\,\circ}_j,\mathbf{x}_j)^2 \right),
\end{equation}
where $e:\mathbb{R}^k \rightarrow \mathbb{R}$ is an error function that satisfies the conditions of identity, symmetry, and triangle inequality.
\textit{The Pareto-optimal point having the minimum \pdu is the selected solution}. The error function $e(\cdot)$ is parametric, i.e., we can set any error or distance metric as $e(\cdot)$, like Euclidean distance or mean squared error.
\end{definition}

\begin{proof}[Derivation]
Let us consider an objective function space $\mathbb{R}^k$, where $k$ is the number of objectives, and a dataset $\mathcal{D}$ of $m$ samples (users/queries). For each sample, we suppose to know the best possible value of each objective. Then, we can associate each sample with a $k$-dimensional vector $\textit{\textbf{f}}^{\,\circ}_j$, with $j \in \{1,\dots,m\}$, which constitutes its generalized utopia point in the objective function space $\mathbb{R}^k$.  We use $\mathbf{F}=\{\textit{\textbf{f}}^{\,\circ}_j\, |\, j\in \{1,\dots,m\}\}$ to denote the set of all the generalized utopia points referring to the $m$ samples.
Let us now consider a 
model $\eta$ that returns $k$ objectives performance values for each sample in $\mathcal{D}$. As before, each sample corresponds to a $k$-dimensional vector $\mathbf{x}_j$, with $j \in \{1,\dots,m\}$, which represents the model performance  for that sample in $\mathbb{R}^k$. We denote $\mathcal{P}=\{\mathbf{x}_j\, |\, j\in \{1,\dots,m\}\}$.
Thus, each sample $j$ is represented by $\textit{\textbf{f}}^{\,\circ}_j$ and $\mathbf{x}_j$ in the objective function space: the closer the points, the better the model $\eta$ performs. Let us introduce an error function $e:\mathbb{R}^k \rightarrow \mathbb{R}$ satisfying the conditions of identity, symmetry,
and triangle inequality. The error of the model $\eta$ on the $j$-th sample is $e(\textit{\textbf{f}}^{\,\circ}_j,\mathbf{x}_j)$.
By supposing the error term follows the IID property, it has a Gaussian distribution with mean $\mu=0$ and variance $\sigma^2$, i.e.,  $e(\textit{\textbf{f}}^{\,\circ}_j, \mathbf{x}_j) \sim \mathcal{N}(0,\sigma^2)$, whose probability density function is:

\begin{equation}\label{eq:pdf}
    p(e(\textit{\textbf{f}}^{\,\circ}_j,\mathbf{x}_j)) = \frac{1}{\sqrt{2\pi}\sigma}\operatorname{exp}\left(-\frac{e(\textit{\textbf{f}}^{\,\circ}_j,\mathbf{x}_j)^2}{2\sigma^2}\right).
\end{equation}

We can note that if $\textit{\textbf{f}}^{\,\circ}_j$ and $\mathbf{x}_j$ are close, the exponent part of Equation \eqref{eq:pdf} tends to $1$, and the probability increases while tending to $0$ when the two points are far apart and the probability decreases.

Then, we compute the error probability density function of the error for the entire dataset $\mathcal{D}$. We observe that the model $\eta$ has 
some parameters $\mathbf{\Theta}$. Hence, $\mathcal{P}$ can be expressed as a function $g$ of the parameters $\mathbf{\Theta}$: $\mathcal{P} = g(\mathbf{\Theta})$. Then, 
a vector $\mathbf{x}_j \in \mathcal{P}$ can be rewritten as $\mathbf{x}_j = g(\mathbf{\Theta})_j$. By assuming the samples to be independent, we obtain the following expression for the likelihood function:

\begin{equation}
    p(e(\mathbf{F},g(\mathbf{\Theta})))=\prod_{j=1}^{m}p(e(\textit{\textbf{f}}^{\,\circ}_j,g(\mathbf{\Theta})_j)).
\end{equation}

Since $\textit{\textbf{f}}^{\,\circ}_j$ is the (generally unattainable) output we desire to have, we are interested in finding the parameters $\mathbf{\Theta}$ for the model $\eta$ such that the likelihood function  $p(e(\mathbf{F},g(\mathbf{\Theta})))$ is the highest. 
As the logarithmic function is increasing monotone, it does not modify the maximum positions. Hence, we can compute the log-likelihood instead of the likelihood to simplify calculations:
\begin{gather}
    \log p(e(\textit{\textbf{f}}^{\,\circ}_j,g(\mathbf{\Theta})))=\log \prod_{j=1}^{m}p(e(\textit{\textbf{f}}^{\,\circ}_j,g(\mathbf{\Theta})_j))\\
    \label{eq:maxlik}
    = m \log \frac{1}{\sqrt{2\pi}\sigma} - \frac{1}{2\sigma^2}\sum_{j=1}^{m} e(\textit{\textbf{f}}^{\,\circ}_j,g(\mathbf{\Theta})_j)^2.
\end{gather}
At this point, we explicit the variance term $\sigma^2$. Since we have supposed that the error term $e(\textit{\textbf{f}}^{\,\circ}_j,\mathbf{x}_j)$ has a Gaussian distribution with $\mu=0$, the variance $\sigma^2$ is defined as $\frac{\sum_{j=1}^{m}e(\textit{\textbf{f}}^{\,\circ}_j,g(\mathbf{\Theta})_j)^2}{m}$. 
By introducing this term in Equation \eqref{eq:maxlik}, we obtain that the log-likelihood is:
\begin{dmath}
    \log p(e(\textit{\textbf{f}}^{\,\circ}_j,g(\mathbf{\Theta}))) = m \log \frac{1}{\sqrt{2\pi}\sqrt{\frac{1}{m}\sum_{j=1}^{m}e(\textit{\textbf{f}}^{\,\circ}_j,g(\mathbf{\Theta})_j)^2}} - \frac{1}{\frac{2}{m}\sum_{j=1}^{m}e(\textit{\textbf{f}}^{\,\circ}_j,g(\mathbf{\Theta})_j)^2}\sum_{j=1}^{m} e(\textit{\textbf{f}}^{\,\circ}_j,g(\mathbf{\Theta})_j)^2  
\end{dmath}
\begin{gather}
    \label{eq:likprov}
    = -m\log(\sqrt{2\pi}) + m\log m -\frac{1}{2}\log \left( \sum_{j=1}^m e(\textit{\textbf{f}}^{\,\circ}_j,g(\mathbf{\Theta})_j)^2 \right) -\frac{m}{2}.
\end{gather}
By supposing to train the model $\eta$ on the same dataset $\mathcal{D}$ with several configurations of $\mathbf{\Theta}$, the terms depending on the dataset size $m$ and the constant $1/2$ in Equation (\ref{eq:likprov}) can be removed as they are constant when choosing the highest log-likelihood. Hence, the only variable quantity among the different log-likelihoods is:
\begin{equation}\label{eq:qua}
     -\log \left( \sum_{j=1}^m e(\textit{\textbf{f}}^{\,\circ}_j,g(\mathbf{\Theta})_j)^2 \right).
\end{equation}
Therefore, we are looking for the model $\eta$ with parameters $\mathbf{\Theta}$ having the maximum value of the term in Equation \eqref{eq:qua}:
\begin{equation}
     \max\left[-\log \left( \sum_{j=1}^m e(\textit{\textbf{f}}^{\,\circ}_j,g(\mathbf{\Theta})_j)^2 \right)\right]. 
\end{equation}
Finally, this remainder term can be easily rewritten with a positive sign as long as we choose the configuration of $\mathbf{\Theta}$ for the model $\eta$ having the minimum value for this quantity:
\begin{equation}
     \min\left[\log \left( \sum_{j=1}^m e(\textit{\textbf{f}}^{\,\circ}_j,g(\mathbf{\Theta})_j)^2 \right)\right] = \min\left[\log \left( \sum_{j=1}^m e(\textit{\textbf{f}}^{\,\circ}_j,\mathbf{x}_j)^2 \right)\right].
\end{equation}
\end{proof}

\subsection{Calibrated \pdu}\label{sez:disccalibrated}
\pdu allows setting a generalized utopia point for each sample of the dataset, i.e., queries and users in an IR or RS scenario, respectively. This feature allows choosing a solution not only based on the ``global''  performance achieved by the IR/RS model, but also in a more fine-grained resolution that now considers multiple quality criteria expressed on a sample level. We call such feature \textit{\textbf{calibration}} since it can be usefully exploited in specific scenarios, e.g., personalization in RS, where it is possible to define generalized utopia points according to individual users' preferences. These generalized utopia points can be fixed apriori, e.g., they can be identified by the system designer or computed through functions that numerically quantify the users' tendencies, similarly to what has been done in previous works regarding \textit{calibrated recommendations}~\cite{DBLP:journals/eswa/JugovacJL17, DBLP:conf/icdm/OhPYSP11, DBLP:conf/recsys/Steck18}. We refer to this feature as \textit{Calibrated}-\pdu (C-\pdu).
\subsection{Feature Comparison}
\label{sec:comparison}
In Section \ref{sec:baseline}, we have presented the most-used techniques to choose a single best solution belonging to a Pareto frontier. However, as also stated by \citet{wu2022multifr}, there is no consensus on the strategy to solve this task in the IR and RS communities. Not surprisingly, all methods have some advantages and limitations, leading to a lack of an ideal strategy~\cite{DBLP:journals/csur/LiY19}. Hence, a comparison of the features provided by \pdu and state-of-the-art techniques is needed. Specifically, we identify some desirable features the techniques should have.
Table~\ref{tab:comparison} discusses the main properties of \pdu and other state-of-the-art techniques.
First, \textbf{the strategy should be suitable even when dealing with more than two objectives}. In this regard, the angle-based knee point is the only ineffective method. 
Second, \textbf{the strategy should not need any additional knowledge}. Most techniques require additional problem information, i.e., the reference point ($\mathcal{HV}$), the (generalized) utopia point (\textit{ED}, \pdu), and a weights set (\textit{WM}). Since the results of a given strategy can  largely depend on such information, a fair strategy should require as less additional information as possible. The weights should be set by a decision-maker with deep knowledge of the hierarchy among the objectives. In contrast, the reference and the (generalized) utopia points are ordinarily intrinsic to the problem. Despite some common practices (e.g., nadir point)~\cite{DBLP:journals/csur/LiY19}, it has been shown that determining a  reference point $\mathbf{r}$ for $\mathcal{HV}$ is generally more challenging~\cite{DBLP:journals/csur/LiY19, hansen1994evaluating}, and a badly defined reference point can lead to inconsistent evaluation results~\cite{DBLP:journals/tec/KnowlesC03}. Indeed, having a reference point slightly different from the nadir point could lead to incongruous evaluation, as experimentally demonstrated by \citet{DBLP:journals/ec/IshibuchiISN18}. Therefore, the utopia point is the most effortlessly additional information that can be exploited for this task.
Third, \textbf{the strategy should not require to scale the range of the objectives}. Scaling may be needed for strategies whose calculation involves objective blending, i.e. \textit{U-KP}, \textit{ED}, \textit{WM}, and \pdu. When the objectives have different scales, the bigger the range of an objective, the bigger its contribution to the selection of a solution. However, the choice of scaling the objectives is left to the system designer. 
Fourth, \textbf{the strategy should be deterministic}. The \textit{U-KP} strategy  requires randomly extracting a set of weights from a uniform distribution. This could potentially affect the consistency and reproducibility of results.
Fifth, \textbf{the strategy should equally promote the solutions despite their position on the Pareto frontier}. The strategies blending the objectives are not biased to select solutions based on particular Pareto frontier regions. This is not true for the $\mathcal{HV}$ strategy that tends to promote the solutions on the concave region of a Pareto frontier.

\begin{table}[t!]
\caption{Overview of the properties of \pdu and other state-of-the-art selection strategies. The symbols \ding{51} (\ding{55},$\textbf{---}$) indicate that the method has (does not have, could not have) the specified property. 
}
\label{tab:comparison}
\footnotesize

\rowcolors{2}{gray!15}{white}
\begin{tabular}{lcccccc}
\toprule
Method & \textit{A-KP} &  \textit{U-KP} & $\mathcal{HV}$& \textit{ED} & \textit{WM} & \pdu \\ \midrule
\begin{tabular}[l]{@{}l@{}}Suitable\\ With \textgreater 2 \\Objectives\end{tabular}            &                \ding{55}                 &                                        \ding{51}                    & \ding{51}                       &         \ding{51}               &     \ding{51}        &  \ding{51}     \\ 
\begin{tabular}[l]{@{}l@{}}No Additional\\Knowledge\\Problem\end{tabular}         &             \ding{51}                    &                                  \ding{51}                        & $\mathbf{r}$                    &  $\textit{\textbf{f}}^{\,\diamond}$                              &    $\mathbf{w}$        &  $\textit{\textbf{f}}^{\,\circ}$                \\ 
\begin{tabular}[l]{@{}l@{}}No Need of\\ Scaling before\\ Calculation\end{tabular}                &       \ding{51}                                  &        $\textbf{---}$        & \ding{51}                       &          $\textbf{---}$             &       $\textbf{---}$          &   $\textbf{---}$  \\ 
Deterministic         &                  \ding{51}               &         \ding{55}                              & \ding{51}                    &           \ding{51}             &       \ding{51}       &   \ding{51}      \\ 
\begin{tabular}[l]{@{}l@{}}Equal\\Treatment of\\PF Regions\end{tabular}                  &                      \ding{55}          &      \ding{51}        &          \ding{55}           &            \ding{51}            &      \ding{51}         &      \ding{51}    \\ 
Calibration            &        \ding{55}                        &       \ding{55}                                     &        \ding{55}           &       \ding{55}              &   \ding{55}     &    \ding{51}   \\ \bottomrule
\end{tabular}
\vspace{-1em}
\end{table}

\vspace{1mm}
\noindent \textbf{Final Observations and Calibration}.
To summarize, none of the strategies own all the properties. 
However, some considerations can be made.
\textit{A-KP} and \textit{U-KP} are characterized by huge drawbacks. The former can be utilized only in contexts considering two objectives. The latter is nondeterministic. 
Furthermore, none of the techniques is able to select a solution irrespective of its position on the Pareto frontier and to be independent of scaling the objective ranges before calculation simultaneously.
In this regard, a system designer could prefer to adopt a technique able to fairly choose a solution despite its position on the Pareto frontier (as done by \textit{U-KP}, \textit{ED}, \textit{WM}, and \pdu). Indeed, scaling the objectives can be easily performed with a simple operation such as min/max normalization. Furthermore, this operation is subject to the system designer, who can consider the objectives range in specific applications. 
Concerning the additional knowledge problem, only \textit{A-KP} and \textit{U-KP} do not need supplementary information. However, as stated before, they are characterized by main drawbacks. Then, such additional knowledge is required. Among the remainder techniques, \pdu and \textit{ED} exploit  easier-to-define additional material, i.e., the utopia point. 

By looking beyond, the proposed \pdu allows us to define a utopia point for each sample in the dataset. While the other approaches exploit only aggregated  models' performance, \pdu  opens to a novel ``calibrated'' way to select one---best Pareto-optimal solution tailored to individual sample characteristics. To the best of our knowledge, this is the first attempt to introduce this kind of feature in the task of Pareto-optimal solutions selection strategy.

From now on, when no confusion arises, we will use \textit{utopia point} to refer also to a \textit{generalized utopia point}.


\section{Experimental Evaluation}
\label{sec:experiments}
We now present an experimental evaluation based on public data that aims at answering the following research questions:
\begin{itemize}
\item[\textbf{RQ1}:] How do \pdu and other state-of-the-art selection strategies behave w.r.t. the discussed properties? (see Section~\ref{sec:comparison})
\item[\textbf{RQ2}:] How does the distribution of the points composing the points on the Pareto frontier influence the selection of a solution?
\item[\textbf{RQ3}:] How does the calibration feature impact the selection of a solution?
\end{itemize}

\subsection{Experimental Scenarios}\label{sec:expsc}
Driven by the observation that, in IR and RS settings, the Pareto frontier is populated by points representing aggregated results, we analyze the selection strategies in these two settings.

\vspace{1mm}
\noindent \textbf{Information Retrieval Scenario}.
\label{sez:expir}
Concerning the IR scenario, we focus on an ad-hoc search task by dealing with the efficiency / effectiveness / energy-consumption trade-off of query processing in  IR systems based on machine-learned ranking models \cite{burges2010ranknet}.
IR systems heavily exploit supervised techniques for learning document ranking models that are both effective and efficient, i.e., able to retrieve within a limited time budget high-quality documents relevant to users' queries. 
State-of-the-art learning-to-rank models include ensembles of regression trees trained with gradient boosting algorithms, e.g., LambdaMART~\cite{lambdamart,burges2010ranknet}, and deep neural networks,  e.g., NeuralNDCG~\cite{DBLP:journals/corr/abs-2102-07831}. Since ranking is a complex task and the training datasets are large, the learned models are complex and computationally expensive at inference time. The tight constraints on query response time thus require suitable solutions to provide an optimal trade-off between efficiency and ranking quality~\cite{SIGIR2015,capannini2016quality,ecir22}.

In  this scenario, we use the LambdaMART \cite{lambdamart,burges2010ranknet} implementation available in LightGBM~\cite{10.5555/3294996.3295074} 
to train ranking models based on ensembles of regression trees and Neural Networks (NN) trained in Pytorch~\cite{DBLP:conf/nips/PaszkeGMLBCKLGA19} following the optimization methodology proposed in \cite{nardini2022distilled}. The models are trained on \msn~\cite{DBLP:journals/corr/QinL13}, a public and widely-used dataset for learning to rank. The evaluation employs $11$ LambdaMART and $5$ Neural Networks ranking models, each tested on the $6$,$306$ queries of the \msn test set.
We measure the ranking quality of each  model in terms of average nDCG@10 $\left(f_1\right)$, and average ranking time (seconds per document) $\left(f_2\right)$. For the LambdaMART configurations, we also measure the average energy consumption (Joules per document) $\left(f_3\right)$. The average ranking time of each model has been measured by using QuickScorer \cite{SIGIR2015}, while  energy consumption has been measured by using the Mammut library \cite{DESENSI2017150}. Efficiency experiments are performed on a dedicated Intel Xeon CPU E5-2630 v3 clocked at 2.4 GHz in single-thread execution. QuickScorer is compiled using GCC 9.2.1 with the \texttt{-O3} option.

\textit{In this IR experimental scenario, we focus on selecting the best efficiency/effectiveness trade-off for query processing}.

\vspace{1mm}
\noindent \textbf{Recommendation Scenario}.
\label{sez:expcalibrated}
Concerning the RS scenario, we consider two of the main problems of recommendation algorithms, i.e., the accuracy of the recommendations and the tendency to over-suggest popular items. Often, the ability of RS to provide accurate recommendations is competing with the capability of including long-tail items in such suggestions~\cite{DBLP:conf/sigir/NaghiaeiRD22}, inducing a trade-off. Hence, we consider two objectives. We compute the Recall $\left(f_1\right)$ to measure the accuracy of suggestions and the average percentage of items in the long-tail (APLT)~\cite{DBLP:conf/recsys/AbdollahpouriBM17} $\left(f_2\right)$ to measure to what extent a RS can recommend unpopular items:
\begin{equation}
    APLT = \frac{1}{|\mathcal{U}_{t}|}\sum_{u \in \mathcal{U}_{t}}\frac{|\{i, i \in (\mathcal{L}_u \cap \Phi)\}|}{|\mathcal{L}_u|},
\end{equation}
where $|\mathcal{U}_{t}|$ is the number of users in the test set, $\mathcal{L}_u$ is the list of recommended items to user $u$, and $\Phi$ is the set of long-tail items. The higher the metric, the higher the number of niche items suggested. 

Specifically, we interpret APLT from two perspectives, identifying two experimental scenarios.
On the one hand, we assess APLT from provider-side fairness. The provider side fairness can be quantified as the models' ability to expose items to users evenly~\cite{DBLP:conf/recsys/AbdollahpouriBM17, DBLP:journals/umuai/AbdollahpouriAB20, wu2022multifr}. Indeed, the over-recommendation of popular items, i.e., the so-called unfairness of popularity bias, may be felt as unfair by providers who get long-tail items under-represented in the suggestions. Hence, in this scenario, the goal is to choose a model that promotes relevant items without affecting niche items' visibility. 

\textit{In this first RS experimental scenario, we focus on selecting the best recommendation model dealing with multiple objectives}.

On the other hand, we evaluate APLT from the final user point of view. Indeed, certain users may prefer to consume popular items, while others niche items. Consequently, exclusively recommending mainstream items would hurt the experience of long-tail users, and vice versa. The approach of calibrated recommendation has shown a valuable solution toward this direction of research~\cite{DBLP:conf/icdm/OhPYSP11, DBLP:conf/recsys/Steck18}. A recommendation list is calibrated concerning popularity when the set of items it covers matches the user’s profile in terms of item popularity~\cite{DBLP:conf/flairs/AbdollahpouriBM19}.
Inspired by the concept of popularity-based calibrated recommendation,  for each user, we compute the values of the APLT target $\left(f_2\right)$ stemming from their popularity profile. To this end, we compute the user-level APLT utopia values using the \textit{weighted combination of mean and standard deviation} method described by \citet{DBLP:journals/eswa/JugovacJL17}. We consider the set of users $\mathcal{U}$, the set of items $\mathcal{I}$, and the mean number of transactions $T$ in the training set. For each item $i\in\mathcal{I}$, we assess its popularity $pop_i$ by counting the number of transactions the item is involved in.  For each user $u\in\mathcal{U}$, we define the set $\Gamma_u=\{pop_i\,|\,u \text{ interacted with } i\}$. We quantify the user $u$ popularity tendencies as
    $pop_u = \alpha \cdot \mu \left(\Gamma_u\right) + \beta \cdot \sigma \left(\Gamma_u\right)$,
where $\alpha$ and $\beta$ are set to a fixed value of $1$ as done in \cite{DBLP:journals/eswa/JugovacJL17}, $\mu\left(\cdot\right)$ and $\sigma\left(\cdot\right)$ are the mean and standard deviation operators, respectively. The higher is $pop_u$, the most user $u$ has consumed mainstream items in her past interactions. 
Finally, we normalize $pop_u$ and compute the APLT utopia value for each user:
\begin{equation}\label{eq:apltutopia}
    f^\circ_2 = h_2(u) = \frac{pop_\Psi - pop_u}{pop_\Psi - pop_\Phi},
\end{equation}
where $\Phi$ and $\Psi$ are the sets composed by $pop_i$ values such that $i$ is one of the less and most $T$ consumed items, respectively. With this normalization, the higher is $f_2^\circ$, the less popular is the user profile.

\textit{In this second RS experimental scenario, we show how important a calibrated technique is for choosing the best recommendation model dealing with multiple objectives}.

In the two experimental scenarios presented for RS, we exploit the \easer recommendation model~\cite{DBLP:conf/www/Steck19}, which works like a shallow autoencoder. This model is characterized by a single hyper-parameter to tune, i.e., the L2-norm regularization ($\lambda$). Nevertheless, it has been shown that it often outperforms other state-of-the-art recommender systems~\cite{DBLP:conf/um/AnelliBNJP22}. Specifically, we explore the hyper-parameter $\lambda$ by training $48$ configurations on the book-domain dataset \textit{Goodreads}~\cite{DBLP:conf/acl/WanMNM19} ($18$,$892$ users, $25$,$475$ items, and $1$,$378$,$033$ transactions) and on the music-domain dataset \textit{Amazon Music}~\cite{DBLP:conf/um/AnelliBNJP22} ($14$,$354$ users, $10$,$027$ items, and $145$,$523$ transactions). We split the datasets following the $70$-$10$-$20$ hold-out strategy. Thus, the evaluation of this scenario employs $48$ solutions on the objective function space, each tested on the remaining users of the test set ($18$,$070$ of \textit{Goodreads}, and $14$,$354$ of \textit{Amazon Music}). 

\input{images/paretofrontscikmms.tex}

\subsection{Experimental Methodology}
The different hyperparameter configurations introduced before, for the two IR and RS settings, generate solutions in the objectives function space for each specific experimental scenario. Once the Pareto-optimal solutions that compose the Pareto frontier are identified, we select one by applying \pdu and the other selection strategies we analyzed in this work. The selected solutions are then analyzed according to the features introduced in Section~\ref{sec:comparison}. Moreover, we investigate in detail how the formulation of \pdu and its calibration feature influence the choice of the one---best solution by looking at the distribution of points composing that solution. 
This Section details the experimental settings employed for each selection strategy. We refer to the reference point and the utopia point with $\mathbf{r}$ and $\textit{\textbf{f}}^{\,\circ}$, respectively. Furthermore, we use the Euclidean distance as $e(\cdot)$ in the formulation of \pdu, to have an immediate comparison with $ED$ to assess the impact of the points distribution composing a solution. Tables~\ref{tab:ir}, \ref{tab:irnn}, and \ref{tab:rs} report the results for the solutions chosen by at least one strategy. For the sake of completeness, the reader may find the complete sets of results in the GitHub repository. The best values for each metric are reported in bold, while the arrows indicate whether better stands for low $\downarrow$ or high $\uparrow$ values.

\vspace{1mm}
\noindent \textbf{Experimental settings for the IR scenario}.
A nadir point cannot be established for the IR scenario because  two of the objectives, i.e., efficiency and energy consumption, are not bounded in the opposite direction of the optimization target. For this reason, we define the reference point by slightly worsening the worst values reached by the optimal solutions available. By doing so, we end up setting $\mathbf{r}=\left(0.5, 0.00002, 0.001\right)$ for $\mathcal{HV}$. Moreover, we set $\textit{\textbf{f}}^{\,\circ}=\left(1,0,0\right)$ for $ED$, and for each sample in the dataset in \pdu. For what regards \textit{WM}, we equally treat the objectives by setting each weight to $0.5$. Finally, in this scenario, we do not apply any normalization to the objective values achieved with the different models.

\vspace{1mm}
\noindent \textbf{Experimental settings for the RS scenario}.
Differently from the IR scenario, a nadir point can be established here because the two objectives under consideration, i.e., \textit{Recall} and \textit{APLT}, are bounded. 
We thus set $\mathbf{r}=\left(0, 0\right)$ for $\mathcal{HV}$, and $\textit{\textbf{f}}^{\,\circ}=\left(1,1\right)$ for $ED$. As before, we give equal importance to the objectives in \textit{WM} by setting each weight to $0$.$5$. Concerning \pdu, we set $f^\circ_1=1$ for each sample utopia point as we want all users to have accurate recommendations. Instead, we set $f^\circ_2=1$ in the first RS experimental scenario, while we compute specific values of $f^\circ_2$ for each user as in Equation~(\ref{eq:apltutopia}) in the second RS experimental scenario. Finally, in both RS scenarios, we apply a min-max normalization to the objectives.
We discuss the results obtained in the different scenarios. We first divide the discussion according to both IR and RS scenarios for RQ1 and RQ2. Then, we answer RQ3 by exploiting the second RS scenario.
\vspace{-0.cm}
\subsection{Performance Comparison (RQ1)}
\noindent \textbf{IR scenario}.
We answer RQ1 by first focusing on the IR scenario. The results for this scenario are summarized in Tables~\ref{tab:ir} (LambdaMART) and~\ref{tab:irnn} (Neural Networks). The plots in Figures~\ref{fig:exp1lambda} and~\ref{fig:exp3dimensions} show the Pareto-optimal points selected by the different techniques for the cases considering two and three objectives regarding the LambdaMART models, respectively. Figure~\ref{fig:exp1nn} shows the points selected in the case of the Neural Networks models. 

Regarding the two-objective case, we observe that the methods blending the objectives (\pdu, \textit{ED}, \textit{WM}) select the same Pareto-optimal solution lying on the boundary of the Pareto frontier for both families of models, thus maximizing the accuracy at the cost of efficiency. 
In contrast, $\mathcal{HV}$ chooses an inner solution of the Pareto frontier in both cases, i.e., more efficient models, 
that however show a significantly lower performance in terms of nDCG compared to the selection provided by \pdu (0.5225 vs. 0.5179 for LambdaMART, and 0.5185 vs. 0.5144 for the Neural Network). 
It is worth noting that, in this case, no transformation has been applied to the scale of the objectives, and the values of the Pareto solutions for what regards the efficiency scale lead the points to be closer to the utopia value $f^{\circ}_2=0$. If a min/max normalization is applied to the objective, \pdu still selects the same solution.
Another essential implication arising from this analysis is that, in this scenario, we cannot establish the nadir point, making challenging the definition of the reference point. Consequently, this potentially leads to different results based on how we define the reference point. Indeed, as we push the reference point away from the Pareto frontier, $\mathcal{HV}$ selects a boundary solution, as done by \pdu.
In light of the above results, we observe that if the information related to the nadir point is unavailable, the definition of the reference point can strongly affect the selection of the final solution.
Moreover, if the reference point is estimated by looking at the collection of the considered solutions, i.e., by slightly increasing the worst values reached by them, $\mathcal{HV}$ promotes the solution in the middle. Indeed, the definition of the reference point in such a way makes the volume of those solutions, computed as in Equation~(\ref{eq:hv}), higher than any other. Thus, $\mathcal{HV}$ unequally considers the remaining points lying on the boundaries of the Pareto frontier. 
Finally, it is worth highlighting that \textit{U-KP}, although reported in Figures~\ref{fig:exp1lambda} and~\ref{fig:exp1nn}, is not deterministic. Indeed, by executing this method several times, it may choose different points as the weights of the utility function (see Section \ref{sec:knee_point}) are randomly extracted from a uniform distribution.

Moving to the three-objective formulation of the IR scenario for the LambdaMART models, Figure~\ref{fig:exp3dimensions} shows that 
when introducing the energy consumption objective, the methods tend to choose a more efficient model than the one selected in the two-objectives scenario. As before, \pdu and \textit{ED} tend to maximize the accuracy with respect to $\mathcal{HV}$ that still select solutions in the middle.
The three-dimensional scenario confirms two behaviors observed in the two-dimensional one. First, the solution selected by $\mathcal{HV}$ depends on the chosen reference point since it is not possible to define a nadir point. Second, \textit{U-KP} still exhibits a non-deterministic behavior.

Finally, we claim that \pdu and \textit{ED} perform the most convenient selection from a qualitative perspective. By looking at Tables~\ref{tab:ir} and~\ref{tab:irnn}, we see that they choose the models with higher values of \textit{nDCG} for all IR cases. Indeed, both efficiency and energy consumption objectives are closer to their respective utopia values. This means that more complex models, chosen by \pdu and \textit{ED}, guarantee considerable improvement in ranking accuracy at a small reduction of efficiency and energy consumption. Conversely, $\mathcal{HV}$ chooses models that exhibit a considerable decrease in terms of nDCG.

\begin{table}
 \caption{Neural Networks selected solutions in the IR scenario. The objectives are accuracy (\textit{nDCG}) and efficiency (\textit{Seconds}).}  
\label{tab:irnn}
\setlength\tabcolsep{1.2pt}
\begin{threeparttable}[para]
\footnotesize
\rowcolors{3}{gray!15}{white}
\begin{tabular}{ccccccccccc}
\toprule
\multicolumn{ 4}{c}{\textbf{Models}} &\multicolumn{ 2}{c}{\textbf{Objectives}}&\multicolumn{ 5}{c}{\textbf{Selection Strategies}} \\
\midrule
      L1 & L2 & L3 & L4    &      nDCG $\uparrow$ &      Seconds $\downarrow$ &      \multicolumn{ 1}{c}{\pdu $\downarrow$} &\multicolumn{ 1}{c}{$\mathcal{HV}\uparrow$} &\multicolumn{ 1}{c}{\textit{U-KP}$\uparrow$} & \multicolumn{ 1}{c}{$ED\downarrow$} & \multicolumn{ 1}{c}{$WM\uparrow$} \\
\midrule
  100&50&50&10 &  0.5144 &  3.3003\tnote{$\dagger$} & 7.5069 & \textbf{2.4099\tnote{$\mathsection$}} &           \textbf{221} &           0.4856 &      0.1286 \\
200&100&100&50 &  \textbf{0.5185} &   1.0476\tnote{*} & \textbf{7.4959} & 1.7598\tnote{$\mathsection$} &           204 &          \textbf{ 0.4815} &      \textbf{0.1296} \\
\bottomrule
\end{tabular}
\begin{tablenotes}\footnotesize
\item[$\dagger$]$\times 10^{-6}$
\item[*]$\times 10^{-5}$ 
\item[$\ddag$]$\times 10^{-8}$ 
\item[$\mathsection$]$\times 10^{-7}$ 
\end{tablenotes}
\vspace{-0.6em}
\end{threeparttable}
\end{table}
{\setlength{\tabcolsep}{1.0pt}
\begin{table}
 \caption{\easer selected solutions (for Goodreads and Amazon Music) in the RS scenario with \textit{Recall} and  \textit{APLT} objectives.} 
\label{tab:rs}
\centering
\footnotesize
\begin{threeparttable}[para]
\rowcolors{3}{white}{gray!15}
\begin{tabular}{ccccccccc}
\toprule
            \multicolumn{ 1}{c}{\textbf{Models}} &\multicolumn{ 2}{c}{\textbf{Objectives}}&\multicolumn{6}{c}{\textbf{Selection Strategies}} \\ \midrule
              \multicolumn{ 1}{c}{$\lambda$} &   \multicolumn{ 1}{c}{Recall $\uparrow$} & \multicolumn{ 1}{c}{APLT $\uparrow$\tnote{*}} &  \multicolumn{ 1}{c}{\pdu$\downarrow$} & \multicolumn{ 1}{c}{C-\pdu$\downarrow$} &  \multicolumn{ 1}{c}{$\mathcal{HV}\uparrow$} & \multicolumn{ 1}{c}{\textit{U-KP}$\uparrow$} & \multicolumn{ 1}{c}{$ED\downarrow$} & \multicolumn{ 1}{c}{$WM\uparrow$} \\
\midrule
\multicolumn{ 9}{c}{\textbf{Goodreads}} \\
\midrule
  0.3 &  0.0384 &  0.0485 & 10.4113 & 10.0898   &   0.1861\tnote{$\dagger$} &     55 &           0.8546 &      \textbf{0.2699} \\
  0.5 &  0.0433 &  0.0443 & \textbf{10.4066} & 10.0829  &   \textbf{0.1919}\tnote{$\dagger$} &    16 &           0.7761 &       0.2686 \\
  1 &  0.0503 &  0.0363 & 10.4098 &  10.0819   &  0.1826\tnote{$\dagger$} &  0 &           \textbf{0.7191} &      0.2546 \\
 60 &  0.0822 &  0.0108 & 10.4126 & \textbf{10.0706}  &   0.0885\tnote{$\dagger$} &   86 &           0.9651 &      0.2556 \\
 90 &  0.0827 &  0.0096 & 10.4134 &  10.0711 &  0.0791\tnote{$\dagger$} &    \textbf{101} &           0.9938 &      0.2510 \\
\midrule
\multicolumn{ 9}{c}{\textbf{Amazon Music}} \\
\midrule
  0.3 &  0.0632 & 0.1976 & \textbf{10.0104}& \textbf{9.8604} &      0.1249\tnote{$\ddag$} &            79 &              0.9524 &         0.2608 \\
  1 &  0.0683 & 0.1898 & 10.0147 & 9.8628 &      \textbf{0.1295}\tnote{$\ddag$} &            49 &              0.8074 &         0.2819 \\
 10 &  0.0853 & 0.1313 & 10.0784 & 9.9160 &      0.1120\tnote{$\ddag$} &             4 &              \textbf{0.6177} &         \textbf{0.2896} \\
 80 &  0.0955 & 0.0766 & 10.1268 & 9.9570 &      0.0731\tnote{$\ddag$} &            \textbf{89} &              0.9780 &         0.2542 \\
\bottomrule
\end{tabular}
\begin{tablenotes}\footnotesize
\item[$\dagger$]$\times 10^{-2}$
\item[$\ddag$]$\times 10^{-1}$ 
\item[*]The higher the better refers to the provider side.
\end{tablenotes}
\end{threeparttable}
\end{table}
}

\vspace{1mm}
\noindent \textbf{RS scenario}.
For  the first RS experimental scenario, we report the results achieved in Table~\ref{tab:rs}  for the Goodreads dataset (Figure~\ref{fig:exp3}) and for the Amazon Music dataset (Figure~\ref{fig:exp3amazon}). 
For both datasets, we notice that two well-separated clusters characterize the Pareto frontier. On the one hand, in Goodreads the \easer configurations with lower L2 norm ($\lambda$) values, which belong to the top-center cluster, account for the accommodation of the objectives. In contrast, the second cluster (bottom-right), i.e.,  $\lambda$ between $10$ and $100$ in Table~\ref{tab:rs}, maximizes \textit{Recall} at the expense of the exposure of the items (lower values of \textit{APLT}). On the other hand, in Amazon Music, these two clusters of configurations follow the opposite behavior. On the one side, the configurations with $\lambda$ between 0.2 and 1 maximize APLT at the detriment of nDCG (top-left cluster). On the other side, the remaining configurations do not promote either \textit{Recall} or \textit{APLT} (bottom-right cluster). In this scenario, $\mathcal{HV}$ suffers less from the problem of promoting solutions in the center of the Pareto frontier. Indeed, differently from the IR scenario, here it is possible to define the nadir point as a reference point because we know the lowest bounds ($0$ for both \textit{APLT} and \textit{Recall}). Consequently, even though $\mathcal{HV}$ selects an inner solution in the Goodreads case, it chooses a point that tends to maximize APLT for the Amazon Music dataset. \pdu follows the behaviour of $\mathcal{HV}$ when selecting the solutions for both datasets. By considering that it selects an outer point of the Pareto frontier in the IR scenario, this endorses the ability of \pdu to equally promote the available solutions despite their positioning on the Pareto frontier. \textit{WM} and \textit{ED} select a solution belonging to the top-center cluster in Goodreads and to the bottom-right cluster in Amazon Music, thus enhancing the trade-off between accuracy measured in terms of \textit{Recall} and items exposure in both cases. Finally, \textit{U-KP} still exhibits a nondeterministic performance. \\
\textit{To  answer RQ1 we conclude observing that \pdu overcomes some limitations of  $\mathcal{HV}$ and \textit{U-KP} competitors. Indeed, \pdu selects one---best---Pareto-optimal solution regardless of its position on the Pareto frontier in a deterministic way. Moreover, 
it exploits the concept of Utopia point as additional information. Such a concept is more convenient to use than the reference point used in $\mathcal{HV}$, since, depending on the problem addressed, the nadir point is difficult to be defined.}

\subsection{Impact of the Points Distribution (RQ2)}
We now answer RQ2 by investigating the impact on selecting the distribution of the points that compose a solution on the Pareto frontier. Indeed, \pdu is the only strategy considering these points in a more fine-grained resolution. This analysis is done on both the IR (Tables~\ref{tab:ir} and~\ref{tab:irnn}) and RS (Table~\ref{tab:rs}) scenarios. To this end, we remember that we have set $e(\cdot)$ as the Euclidean Distance in the formulation of \pdu (Equation (\ref{eq:pdu})). Hence, even if both \pdu and \textit{ED} rely on the Euclidean distance, they work differently in the two experimental scenarios. This observation provides insights on the impact of the points distribution on the selection.

\vspace{1mm}
\noindent \textbf{RS scenario}. \pdu and \textit{ED} choose different solutions for both RS datasets. In this regard, the user data points' distribution in the objective function space plays a crucial role, as visually depicted by Figure~\ref{fig:easerpdued} for the Goodreads dataset. Indeed, the distribution of the solution with $\lambda=0.5$, chosen by \pdu, shows that more points are oriented to the Utopia point than the ones of the solution selected by \textit{ED}. To confirm this fact, we compute the users points' mean Euclidean distances to the utopia point of both solutions. Results confirm that the \easer configuration selected by \pdu has a lower value of average Euclidean distance, i.e., $1$.$3498$ for $\lambda=0.5$, w.r.t. the configuration chosen by \textit{ED}, i.e., $1$.$352$ for $\lambda=1$. The same impact is observed regarding the Amazon Music dataset. Here, \pdu and $ED$ select different configuration models having $\lambda=0.3$ and $\lambda=10$, respectively. As before, the \easer configuration selected by \pdu ($\lambda=0.3$) has a lower value of average Euclidean distance, i.e., $1$.$2361$ than the configuration chosen by \textit{ED} ($\lambda=10$), i.e., $1$.$279$.

\vspace{1mm}
\noindent \textbf{IR scenario}. Concerning the IR two-objectives cases, \pdu and $ED$ choose the same solution for both LambdaMART and Neural Networks models. When introducing energy consumption as the third objective for the LambdaMART models, $ED$ still selects the same configuration with 878 trees and 64 leaves. Conversely, \pdu chooses a more efficient model (300 trees and 64 leaves). Once more, the query points' mean Euclidean distances to the common utopia point of the model selected by \pdu are lower than the ones of the model chosen by $ED$ (0.4813 vs. 0.4945).

\textit{To conclude, the answer to RQ2 is that the distribution of the points composing a solution with respect to a common utopia point has a significant impact on the final selection. This is an important fact, as it paves the way to defining selection strategies that take the distribution of the points into account while performing a selection that can be done in a more---fine-grained---sample-level way.}

\subsection{Impact of Calibration on the Selection (RQ3)}
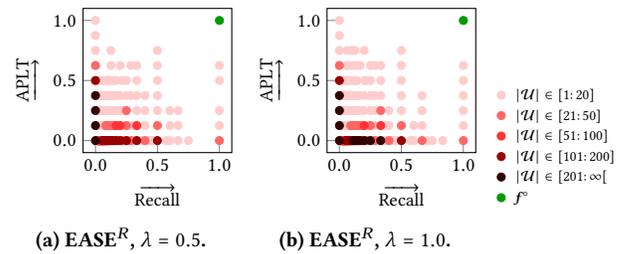
\begin{figure}
\footnotesize
    \centering
   
     \subfloat[\easer, $\lambda=0.5$.]{
    \begin{tikzpicture}
        \begin{axis}[
        clip mode=individual,
        every axis legend/.code={\let\addlegendentry\relax} ,
        width=0.2*(\textwidth/\linewidth)*\linewidth,
        height=3.5cm,
        xlabel=$\ra{\text{Recall}}$,
        ylabel=$\underrightarrow{\text{APLT}}$,
        ylabel near ticks,
        xlabel near ticks,
        xtick pos=left,
        ytick pos=left,
        scaled x ticks=false,
        scaled y ticks=false,
        yticklabel style={
        /pgf/number format/.cd,
        fixed, fixed zerofill,
        precision=1
        },
        xticklabel style={
        /pgf/number format/.cd,
        fixed, fixed zerofill,
        precision=1
        }
        ]

         \addplot[ylgnbu1, only marks, mark=*, mark options={fill=ylgnbu1}, mark size=1.5pt] table [
        x expr={(\thisrow{cat}==0?\thisrow{Recall}:nan},
        y=APLT, col sep=comma]{images/csv/easer05_cat.csv};

        \addplot[ylgnbu2, only marks, mark=*, mark options={fill=ylgnbu2}, mark size=1.5pt] table [
        x expr={(\thisrow{cat}==1?\thisrow{Recall}:nan},
        y=APLT, col sep=comma]{images/csv/easer05_cat.csv};

        \addplot[ylgnbu3, only marks, mark=*, mark options={fill=ylgnbu3}, mark size=1.5pt] table [
        x expr={(\thisrow{cat}==2?\thisrow{Recall}:nan},
        y=APLT, col sep=comma]{images/csv/easer05_cat.csv};

 \addplot[ylgnbu4, only marks, mark=*, mark options={fill=ylgnbu4}, mark size=1.5pt] table [
        x expr={(\thisrow{cat}==3?\thisrow{Recall}:nan},
        y=APLT, col sep=comma]{images/csv/easer05_cat.csv};
        
        \addplot[ylgnbu5, only marks, mark=*, mark options={fill=ylgnbu5}, mark size=1.5pt] table [
        x expr={(\thisrow{cat}==4?\thisrow{Recall}:nan},
        y=APLT, col sep=comma]{images/csv/easer05_cat.csv};

        \addplot[ylgnbu6, only marks, mark=*, mark options={fill=ylgnbu6}, mark size=1.5pt] table [
        x=x,
        y=y, col sep=comma]{images/csv/utopia.csv};

        \end{axis}
        \label{fig:easer05}
    \end{tikzpicture}
    }
    \,
     \subfloat[\easer, $\lambda=1.0$.]{
    \begin{tikzpicture}
        \begin{axis}[
        clip mode=individual,
        every axis legend/.code={\let\addlegendentry\relax} ,
        width=0.2*(\textwidth/\linewidth)*\linewidth,
        height=3.5cm,
        xlabel=$\ra{\text{Recall}}$,
        ylabel=$\underrightarrow{\text{APLT}}$,
        ylabel near ticks,
        xlabel near ticks,
        xtick pos=left,
        ytick pos=left,
        scaled x ticks=false,
        scaled y ticks=false,
        yticklabel style={
        /pgf/number format/.cd,
        fixed, fixed zerofill,
        precision=1
        },
        xticklabel style={
        /pgf/number format/.cd,
        fixed, fixed zerofill,
        precision=1
        }
        ]
         
         \addplot[ylgnbu1, only marks, mark=*, mark options={fill=ylgnbu1}, mark size=1.5pt] table [
        x expr={(\thisrow{cat}==0?\thisrow{Recall}:nan},
        y=APLT, col sep=comma]{images/csv/easer10_cat.csv};

        \addplot[ylgnbu2, only marks, mark=*, mark options={fill=ylgnbu2}, mark size=1.5pt] table [
        x expr={(\thisrow{cat}==1?\thisrow{Recall}:nan},
        y=APLT, col sep=comma]{images/csv/easer10_cat.csv};

        \addplot[ylgnbu3, only marks, mark=*, mark options={fill=ylgnbu3}, mark size=1.5pt] table [
        x expr={(\thisrow{cat}==2?\thisrow{Recall}:nan},
        y=APLT, col sep=comma]{images/csv/easer10_cat.csv};

 \addplot[ylgnbu4, only marks, mark=*, mark options={fill=ylgnbu4}, mark size=1.5pt] table [
        x expr={(\thisrow{cat}==3?\thisrow{Recall}:nan},
        y=APLT, col sep=comma]{images/csv/easer10_cat.csv};
        
        \addplot[ylgnbu5, only marks, mark=*, mark options={fill=ylgnbu5}, mark size=1.5pt] table [
        x expr={(\thisrow{cat}==4?\thisrow{Recall}:nan},
        y=APLT, col sep=comma]{images/csv/easer10_cat.csv};

           \addplot[ylgnbu6, only marks, mark=*, mark options={fill=ylgnbu6}, mark size=1.5pt] table [
        x=x,
        y=y, col sep=comma]{images/csv/utopia.csv};
        \end{axis}
        \label{fig:easer10}
    \end{tikzpicture}
    }\,
    \begin{adjustbox}{width=0.1\textwidth}
    \begin{tikzpicture}
        \begin{customlegend}[        legend columns=1, legend style={draw=none,column sep=1ex}, legend cell align={left},
        legend entries={$\lvert\mathcal{U}\rvert \in [1\colon 20]$, $\lvert\mathcal{U}\rvert \in [21\colon 50]$, $\lvert\mathcal{U}\rvert \in [51\colon 100]$, $\lvert\mathcal{U}\rvert \in [101\colon 200]$, $\lvert\mathcal{U}\rvert \in [201\colon \infty[$, $\textit{\textbf{f}}^{\,\circ}$}]
        \addlegendimage{only marks, color=ylgnbu1,mark=*}
         \addlegendimage{only marks, color=ylgnbu2,mark=*}
        \addlegendimage{only marks, color=ylgnbu3,mark=*}
        \addlegendimage{only marks, color=ylgnbu4,mark=*}
        \addlegendimage{only marks, color=ylgnbu5,mark=*}
        \addlegendimage{only marks, color=ylgnbu6,mark=*}
        \end{customlegend}
    \end{tikzpicture}
    \end{adjustbox}
    \vspace{-0.5em}
    \caption{Distribution of users data points in the objective function space \textit{Recall} / \textit{APLT} for the solutions selected by \pdu (left) and \textit{ED} (right). The color of the point indicates the number of users in the point.\vspace{-3mm}}
\label{fig:easerpdued}
\end{figure}
Finally, we analyze the impact of the calibration introduced for \pdu using the second RS scenario, where we aim to tailor the selection according to the users' item popularity tastes. To this end, we assess the selection performed by Calibrated-\pdu (C-\pdu). Starting from the Amazon Music dataset, the average of the APLT utopia values computed with Equation (\ref{eq:apltutopia}) (0.83) reveals that the dataset's users generally prefer less popular items. Indeed, C-\pdu selects the \easer model with $\lambda=0.3$. This solution lies on the top-left cluster of Figure\ref{fig:exp3amazon}, by maximizing APLT with a loss of Recall. In this case, C-\pdu behaves similarly to \pdu and $\mathcal{HV}$. Moving to the Goodreads dataset, it is characterized by users with more mainstream tastes, since the average of the APLT utopia values is equal to 0.65. Surprisingly, C-\pdu is the only strategy among the ones tested selecting a model configuration belonging to the bottom-right cluster in Figure~\ref{fig:easerpdued} where the solutions achieve higher accuracy values without promoting APLT and following the mainstream users tastes --- along with \textit{U-KP} that, however, has a non-deterministic behavior. These experimental results already qualitatively show the impact of defining a utopia point for each user on the final selection, since C-\pdu is the only strategy to capture the users' popularity profiles for both datasets. We deepen the analysis further by considering the model configurations chosen by \pdu and C-\pdu for Goodreads, i.e., $\lambda=0.5$ and $\lambda=60$, respectively. We observe that, although the model with $\lambda=0.5$ performs better on average APLT, the model with $\lambda=60$ has a lower variance of the mean absolute error ($0$.$036$ for $\lambda=60$ vs. $0$.$039$ for $\lambda=0.5$) between the utopia values and the model performance values for each user.
This indicates that C-\pdu selects the model that generally follows better the users' popularity profile. In addition, this model provides more accurate recommendations on average. Hence,  C-\pdu chooses the model that performs better in terms of accuracy and also tailors the popular tastes of the users.

\textit{To conclude, the answer to RQ3 is that the calibration feature of \pdu allows dealing with the ideal targets for each sample. This confirms that calibration is a viable way to move the selection of the Pareto-optimal solution to a more fine-grained resolution that can lead to significantly different choices in terms of the trade-off selected.}

\section{Conclusion and Future Work}
\label{sec:conclusions}
In this work, we proposed \pdu, a novel, theoretically-justified \emph{post-hoc} technique to select one---best---Pareto-optimal solution among the ones lying in the Pareto frontier in search and recommendation scenarios. 
To our knowledge, \pdu is the only selection technique in the literature that can be ``calibrated'', i.e., it can choose the best Pareto-optimal solution based on ideal targets expressed on single queries or users. We comprehensively compared the properties of \pdu with those of competitor techniques. We conducted an extensive experimental evaluation focusing on both IR and RS scenarios, showing that the formulation and the calibration feature of \pdu have a notable impact on the solution's selection. In the future, we will explore \pdu by exploiting other distance metrics (e.g., Chebyshev and Manhattan). This work could open to the formulation of a new loss function based on the \pdu derivation, to directly train a ranking model on multiple objectives simultaneously.



\bibliographystyle{ACM-Reference-Format}
\bibliography{sample-base}


\begin{thebibliography}{54}


\ifx \showCODEN    \undefined \def \showCODEN     #1{\unskip}     \fi
\ifx \showDOI      \undefined \def \showDOI       #1{#1}\fi
\ifx \showISBNx    \undefined \def \showISBNx     #1{\unskip}     \fi
\ifx \showISBNxiii \undefined \def \showISBNxiii  #1{\unskip}     \fi
\ifx \showISSN     \undefined \def \showISSN      #1{\unskip}     \fi
\ifx \showLCCN     \undefined \def \showLCCN      #1{\unskip}     \fi
\ifx \shownote     \undefined \def \shownote      #1{#1}          \fi
\ifx \showarticletitle \undefined \def \showarticletitle #1{#1}   \fi
\ifx \showURL      \undefined \def \showURL       {\relax}        \fi
\providecommand\bibfield[2]{#2}
\providecommand\bibinfo[2]{#2}
\providecommand\natexlab[1]{#1}
\providecommand\showeprint[2][]{arXiv:#2}

\bibitem[Abdollahpouri et~al\mbox{.}(2020)]%
        {DBLP:journals/umuai/AbdollahpouriAB20}
\bibfield{author}{\bibinfo{person}{Himan Abdollahpouri},
  \bibinfo{person}{Gediminas Adomavicius}, \bibinfo{person}{Robin Burke},
  \bibinfo{person}{Ido Guy}, \bibinfo{person}{Dietmar Jannach},
  \bibinfo{person}{Toshihiro Kamishima}, \bibinfo{person}{Jan Krasnodebski},
  {and} \bibinfo{person}{Luiz~Augusto Pizzato}.}
  \bibinfo{year}{2020}\natexlab{}.
\newblock \showarticletitle{Multistakeholder recommendation: Survey and
  research directions}.
\newblock \bibinfo{journal}{\emph{User Model. User Adapt. Interact.}}
  \bibinfo{volume}{30}, \bibinfo{number}{1} (\bibinfo{year}{2020}),
  \bibinfo{pages}{127--158}.
\newblock
\urldef\tempurl%
\url{https://doi.org/10.1007/s11257-019-09256-1}
\showDOI{\tempurl}


\bibitem[Abdollahpouri et~al\mbox{.}(2017)]%
        {DBLP:conf/recsys/AbdollahpouriBM17}
\bibfield{author}{\bibinfo{person}{Himan Abdollahpouri}, \bibinfo{person}{Robin
  Burke}, {and} \bibinfo{person}{Bamshad Mobasher}.}
  \bibinfo{year}{2017}\natexlab{}.
\newblock \showarticletitle{Controlling Popularity Bias in Learning-to-Rank
  Recommendation}. In \bibinfo{booktitle}{\emph{Proceedings of the Eleventh
  {ACM} Conference on Recommender Systems, RecSys 2017, Como, Italy, August
  27-31, 2017}}, \bibfield{editor}{\bibinfo{person}{Paolo Cremonesi},
  \bibinfo{person}{Francesco Ricci}, \bibinfo{person}{Shlomo Berkovsky}, {and}
  \bibinfo{person}{Alexander Tuzhilin}} (Eds.). \bibinfo{publisher}{{ACM}},
  \bibinfo{pages}{42--46}.
\newblock
\urldef\tempurl%
\url{https://doi.org/10.1145/3109859.3109912}
\showDOI{\tempurl}


\bibitem[Abdollahpouri et~al\mbox{.}(2019)]%
        {DBLP:conf/flairs/AbdollahpouriBM19}
\bibfield{author}{\bibinfo{person}{Himan Abdollahpouri}, \bibinfo{person}{Robin
  Burke}, {and} \bibinfo{person}{Bamshad Mobasher}.}
  \bibinfo{year}{2019}\natexlab{}.
\newblock \showarticletitle{Managing Popularity Bias in Recommender Systems
  with Personalized Re-Ranking}. In \bibinfo{booktitle}{\emph{Proceedings of
  the Thirty-Second International Florida Artificial Intelligence Research
  Society Conference, Sarasota, Florida, USA, May 19-22 2019}},
  \bibfield{editor}{\bibinfo{person}{Roman Bart{\'{a}}k} {and}
  \bibinfo{person}{Keith~W. Brawner}} (Eds.). \bibinfo{publisher}{{AAAI}
  Press}, \bibinfo{pages}{413--418}.
\newblock
\urldef\tempurl%
\url{https://aaai.org/ocs/index.php/FLAIRS/FLAIRS19/paper/view/18199}
\showURL{%
\tempurl}


\bibitem[Anelli et~al\mbox{.}(2022)]%
        {DBLP:conf/um/AnelliBNJP22}
\bibfield{author}{\bibinfo{person}{Vito~Walter Anelli},
  \bibinfo{person}{Alejandro Bellog{\'{\i}}n}, \bibinfo{person}{Tommaso~Di
  Noia}, \bibinfo{person}{Dietmar Jannach}, {and} \bibinfo{person}{Claudio
  Pomo}.} \bibinfo{year}{2022}\natexlab{}.
\newblock \showarticletitle{Top-N Recommendation Algorithms: {A} Quest for the
  State-of-the-Art}. In \bibinfo{booktitle}{\emph{{UMAP} '22: 30th {ACM}
  Conference on User Modeling, Adaptation and Personalization, Barcelona,
  Spain, July 4 - 7, 2022}}, \bibfield{editor}{\bibinfo{person}{Alejandro
  Bellog{\'{\i}}n}, \bibinfo{person}{Ludovico Boratto},
  \bibinfo{person}{Olga~C. Santos}, \bibinfo{person}{Liliana Ardissono}, {and}
  \bibinfo{person}{Bart~P. Knijnenburg}} (Eds.). \bibinfo{publisher}{{ACM}},
  \bibinfo{pages}{121--131}.
\newblock
\urldef\tempurl%
\url{https://doi.org/10.1145/3503252.3531292}
\showDOI{\tempurl}


\bibitem[Branke et~al\mbox{.}(2004)]%
        {DBLP:conf/ppsn/BrankeDDO04}
\bibfield{author}{\bibinfo{person}{J{\"{u}}rgen Branke},
  \bibinfo{person}{Kalyanmoy Deb}, \bibinfo{person}{Henning Dierolf}, {and}
  \bibinfo{person}{Matthias Osswald}.} \bibinfo{year}{2004}\natexlab{}.
\newblock \showarticletitle{Finding Knees in Multi-objective Optimization}. In
  \bibinfo{booktitle}{\emph{Parallel Problem Solving from Nature - {PPSN} VIII,
  8th International Conference, Birmingham, UK, September 18-22, 2004,
  Proceedings}} \emph{(\bibinfo{series}{Lecture Notes in Computer Science},
  Vol.~\bibinfo{volume}{3242})}, \bibfield{editor}{\bibinfo{person}{Xin Yao},
  \bibinfo{person}{Edmund~K. Burke}, \bibinfo{person}{Jos{\'{e}}~Antonio
  Lozano}, \bibinfo{person}{Jim Smith}, \bibinfo{person}{Juan
  Juli{\'{a}}n~Merelo Guerv{\'{o}}s}, \bibinfo{person}{John~A. Bullinaria},
  \bibinfo{person}{Jonathan~E. Rowe}, \bibinfo{person}{Peter Ti{\~{n}}o},
  \bibinfo{person}{Ata Kab{\'{a}}n}, {and} \bibinfo{person}{Hans{-}Paul
  Schwefel}} (Eds.). \bibinfo{publisher}{Springer}, \bibinfo{pages}{722--731}.
\newblock
\urldef\tempurl%
\url{https://doi.org/10.1007/978-3-540-30217-9\_73}
\showDOI{\tempurl}


\bibitem[Branke et~al\mbox{.}(2008)]%
        {DBLP:conf/dagstuhl/2008moo}
\bibfield{editor}{\bibinfo{person}{J{\"{u}}rgen Branke},
  \bibinfo{person}{Kalyanmoy Deb}, \bibinfo{person}{Kaisa Miettinen}, {and}
  \bibinfo{person}{Roman Slowinski}} (Eds.). \bibinfo{year}{2008}\natexlab{}.
\newblock \bibinfo{booktitle}{\emph{Multiobjective Optimization, Interactive
  and Evolutionary Approaches [outcome of Dagstuhl seminars]}}.
  \bibinfo{series}{Lecture Notes in Computer Science},
  Vol.~\bibinfo{volume}{5252}. \bibinfo{publisher}{Springer}.
\newblock
\showISBNx{978-3-540-88907-6}
\urldef\tempurl%
\url{https://doi.org/10.1007/978-3-540-88908-3}
\showDOI{\tempurl}


\bibitem[Burges(2010)]%
        {burges2010ranknet}
\bibfield{author}{\bibinfo{person}{Christopher~JC Burges}.}
  \bibinfo{year}{2010}\natexlab{}.
\newblock \showarticletitle{From ranknet to lambdarank to lambdamart: An
  overview}.
\newblock \bibinfo{journal}{\emph{Learning}} \bibinfo{volume}{11},
  \bibinfo{number}{23-581} (\bibinfo{year}{2010}), \bibinfo{pages}{81}.
\newblock


\bibitem[Capannini et~al\mbox{.}(2016)]%
        {capannini2016quality}
\bibfield{author}{\bibinfo{person}{Gabriele Capannini},
  \bibinfo{person}{Claudio Lucchese}, \bibinfo{person}{Franco~Maria Nardini},
  \bibinfo{person}{Salvatore Orlando}, \bibinfo{person}{Raffaele Perego}, {and}
  \bibinfo{person}{Nicola Tonellotto}.} \bibinfo{year}{2016}\natexlab{}.
\newblock \showarticletitle{Quality Versus Efficiency in Document Scoring with
  Learning-to-rank Models}.
\newblock \bibinfo{journal}{\emph{Information Processing Management}}
  \bibinfo{volume}{52}, \bibinfo{number}{6} (\bibinfo{date}{Nov.}
  \bibinfo{year}{2016}), \bibinfo{pages}{1161--1177}.
\newblock
\showISSN{0306-4573}


\bibitem[Carmel et~al\mbox{.}(2020)]%
        {DBLP:conf/www/CarmelHLL20}
\bibfield{author}{\bibinfo{person}{David Carmel}, \bibinfo{person}{Elad
  Haramaty}, \bibinfo{person}{Arnon Lazerson}, {and} \bibinfo{person}{Liane
  Lewin{-}Eytan}.} \bibinfo{year}{2020}\natexlab{}.
\newblock \showarticletitle{Multi-Objective Ranking Optimization for Product
  Search Using Stochastic Label Aggregation}. In
  \bibinfo{booktitle}{\emph{{WWW} '20: The Web Conference 2020, Taipei, Taiwan,
  April 20-24, 2020}}, \bibfield{editor}{\bibinfo{person}{Yennun Huang},
  \bibinfo{person}{Irwin King}, \bibinfo{person}{Tie{-}Yan Liu}, {and}
  \bibinfo{person}{Maarten van Steen}} (Eds.). \bibinfo{publisher}{{ACM} /
  {IW3C2}}, \bibinfo{pages}{373--383}.
\newblock
\urldef\tempurl%
\url{https://doi.org/10.1145/3366423.3380122}
\showDOI{\tempurl}


\bibitem[Dai et~al\mbox{.}(2011)]%
        {dai2011learning}
\bibfield{author}{\bibinfo{person}{Na Dai}, \bibinfo{person}{Milad Shokouhi},
  {and} \bibinfo{person}{Brian~D Davison}.} \bibinfo{year}{2011}\natexlab{}.
\newblock \showarticletitle{Learning to rank for freshness and relevance}. In
  \bibinfo{booktitle}{\emph{Proceedings of the 34th international ACM SIGIR
  conference on Research and development in Information Retrieval}}.
  \bibinfo{pages}{95--104}.
\newblock


\bibitem[Dalal et~al\mbox{.}(2012)]%
        {10.1145/2187836.2187894}
\bibfield{author}{\bibinfo{person}{Onkar Dalal}, \bibinfo{person}{Srinivasan~H.
  Sengemedu}, {and} \bibinfo{person}{Subhajit Sanyal}.}
  \bibinfo{year}{2012}\natexlab{}.
\newblock \showarticletitle{Multi-Objective Ranking of Comments on Web}. In
  \bibinfo{booktitle}{\emph{Proceedings of the 21st International Conference on
  World Wide Web}} (Lyon, France) \emph{(\bibinfo{series}{WWW '12})}.
  \bibinfo{publisher}{Association for Computing Machinery},
  \bibinfo{address}{New York, NY, USA}, \bibinfo{pages}{419–428}.
\newblock
\showISBNx{9781450312295}
\urldef\tempurl%
\url{https://doi.org/10.1145/2187836.2187894}
\showDOI{\tempurl}


\bibitem[Deb and Gupta(2011)]%
        {doi:10.1080/0305215X.2010.548863}
\bibfield{author}{\bibinfo{person}{Kalyanmoy Deb} {and} \bibinfo{person}{Shivam
  Gupta}.} \bibinfo{year}{2011}\natexlab{}.
\newblock \showarticletitle{Understanding knee points in bicriteria problems
  and their implications as preferred solution principles}.
\newblock \bibinfo{journal}{\emph{Engineering Optimization}}
  \bibinfo{volume}{43}, \bibinfo{number}{11} (\bibinfo{year}{2011}),
  \bibinfo{pages}{1175--1204}.
\newblock
\urldef\tempurl%
\url{https://doi.org/10.1080/0305215X.2010.548863}
\showDOI{\tempurl}
\showeprint{https://doi.org/10.1080/0305215X.2010.548863}


\bibitem[De Sensi et~al\mbox{.}(2017)]%
        {DESENSI2017150}
\bibfield{author}{\bibinfo{person}{Daniele De Sensi}, \bibinfo{person}{Massimo
  Torquati}, {and} \bibinfo{person}{Marco Danelutto}.}
  \bibinfo{year}{2017}\natexlab{}.
\newblock \showarticletitle{Mammut: High-level management of system knobs and
  sensors}.
\newblock \bibinfo{journal}{\emph{SoftwareX}}  \bibinfo{volume}{6}
  (\bibinfo{year}{2017}), \bibinfo{pages}{150--154}.
\newblock
\showISSN{2352-7110}
\urldef\tempurl%
\url{https://doi.org/10.1016/j.softx.2017.06.005}
\showDOI{\tempurl}


\bibitem[Fleischer(2003)]%
        {DBLP:conf/emo/Fleischer03}
\bibfield{author}{\bibinfo{person}{M. Fleischer}.}
  \bibinfo{year}{2003}\natexlab{}.
\newblock \showarticletitle{The Measure of Pareto Optima}. In
  \bibinfo{booktitle}{\emph{Evolutionary Multi-Criterion Optimization, Second
  International Conference, {EMO} 2003, Faro, Portugal, April 8-11, 2003,
  Proceedings}} \emph{(\bibinfo{series}{Lecture Notes in Computer Science},
  Vol.~\bibinfo{volume}{2632})}, \bibfield{editor}{\bibinfo{person}{Carlos~M.
  Fonseca}, \bibinfo{person}{Peter~J. Fleming}, \bibinfo{person}{Eckart
  Zitzler}, \bibinfo{person}{Kalyanmoy Deb}, {and} \bibinfo{person}{Lothar
  Thiele}} (Eds.). \bibinfo{publisher}{Springer}, \bibinfo{pages}{519--533}.
\newblock
\urldef\tempurl%
\url{https://doi.org/10.1007/3-540-36970-8\_37}
\showDOI{\tempurl}


\bibitem[Ge et~al\mbox{.}(2022)]%
        {DBLP:conf/wsdm/GeZYPHHZ22}
\bibfield{author}{\bibinfo{person}{Yingqiang Ge}, \bibinfo{person}{Xiaoting
  Zhao}, \bibinfo{person}{Lucia Yu}, \bibinfo{person}{Saurabh Paul},
  \bibinfo{person}{Diane Hu}, \bibinfo{person}{Chu{-}Cheng Hsieh}, {and}
  \bibinfo{person}{Yongfeng Zhang}.} \bibinfo{year}{2022}\natexlab{}.
\newblock \showarticletitle{Toward Pareto Efficient Fairness-Utility Trade-off
  in Recommendation through Reinforcement Learning}. In
  \bibinfo{booktitle}{\emph{{WSDM} '22: The Fifteenth {ACM} International
  Conference on Web Search and Data Mining, Virtual Event / Tempe, AZ, USA,
  February 21 - 25, 2022}}, \bibfield{editor}{\bibinfo{person}{K.~Selcuk
  Candan}, \bibinfo{person}{Huan Liu}, \bibinfo{person}{Leman Akoglu},
  \bibinfo{person}{Xin~Luna Dong}, {and} \bibinfo{person}{Jiliang Tang}}
  (Eds.). \bibinfo{publisher}{{ACM}}, \bibinfo{pages}{316--324}.
\newblock
\urldef\tempurl%
\url{https://doi.org/10.1145/3488560.3498487}
\showDOI{\tempurl}


\bibitem[Gil{-}Costa et~al\mbox{.}(2022)]%
        {ecir22}
\bibfield{author}{\bibinfo{person}{Veronica Gil{-}Costa},
  \bibinfo{person}{Fernando Loor}, \bibinfo{person}{Romina Molina},
  \bibinfo{person}{Franco~Maria Nardini}, \bibinfo{person}{Raffaele Perego},
  {and} \bibinfo{person}{Salvatore Trani}.} \bibinfo{year}{2022}\natexlab{}.
\newblock \showarticletitle{Ensemble Model Compression for Fast and
  Energy-Efficient Ranking on FPGAs}. In \bibinfo{booktitle}{\emph{Advances in
  Information Retrieval - 44th European Conference on {IR} Research, {ECIR}
  2022, Stavanger, Norway, April 10-14, 2022, Proceedings, Part {I}}}
  \emph{(\bibinfo{series}{Lecture Notes in Computer Science},
  Vol.~\bibinfo{volume}{13185})}, \bibfield{editor}{\bibinfo{person}{Matthias
  Hagen}, \bibinfo{person}{Suzan Verberne}, \bibinfo{person}{Craig Macdonald},
  \bibinfo{person}{Christin Seifert}, \bibinfo{person}{Krisztian Balog},
  \bibinfo{person}{Kjetil N{\o}rv{\aa}g}, {and} \bibinfo{person}{Vinay Setty}}
  (Eds.). \bibinfo{publisher}{Springer}, \bibinfo{pages}{260--273}.
\newblock
\urldef\tempurl%
\url{https://doi.org/10.1007/978-3-030-99736-6\_18}
\showDOI{\tempurl}


\bibitem[Haimes et~al\mbox{.}(1971)]%
        {4308298}
\bibfield{author}{\bibinfo{person}{Haimes}, \bibinfo{person}{Lasdon}, {and}
  \bibinfo{person}{Wismer}.} \bibinfo{year}{1971}\natexlab{}.
\newblock \showarticletitle{On a Bicriterion Formulation of the Problems of
  Integrated System Identification and System Optimization}.
\newblock \bibinfo{journal}{\emph{IEEE Transactions on Systems, Man, and
  Cybernetics}} \bibinfo{volume}{SMC-1}, \bibinfo{number}{3}
  (\bibinfo{year}{1971}), \bibinfo{pages}{296--297}.
\newblock
\urldef\tempurl%
\url{https://doi.org/10.1109/TSMC.1971.4308298}
\showDOI{\tempurl}


\bibitem[Hansen and Jaszkiewicz(1994)]%
        {hansen1994evaluating}
\bibfield{author}{\bibinfo{person}{Michael~Pilegaard Hansen} {and}
  \bibinfo{person}{Andrzej Jaszkiewicz}.} \bibinfo{year}{1994}\natexlab{}.
\newblock \bibinfo{booktitle}{\emph{Evaluating the quality of approximations to
  the non-dominated set}}.
\newblock \bibinfo{publisher}{IMM, Department of Mathematical Modelling,
  Technical Universityof Denmark}.
\newblock


\bibitem[Ishibuchi et~al\mbox{.}(2018)]%
        {DBLP:journals/ec/IshibuchiISN18}
\bibfield{author}{\bibinfo{person}{Hisao Ishibuchi}, \bibinfo{person}{Ryo
  Imada}, \bibinfo{person}{Yu Setoguchi}, {and} \bibinfo{person}{Yusuke
  Nojima}.} \bibinfo{year}{2018}\natexlab{}.
\newblock \showarticletitle{How to Specify a Reference Point in Hypervolume
  Calculation for Fair Performance Comparison}.
\newblock \bibinfo{journal}{\emph{Evol. Comput.}} \bibinfo{volume}{26},
  \bibinfo{number}{3} (\bibinfo{year}{2018}).
\newblock
\urldef\tempurl%
\url{https://doi.org/10.1162/evco\_a\_00226}
\showDOI{\tempurl}


\bibitem[Jugovac et~al\mbox{.}(2017)]%
        {DBLP:journals/eswa/JugovacJL17}
\bibfield{author}{\bibinfo{person}{Michael Jugovac}, \bibinfo{person}{Dietmar
  Jannach}, {and} \bibinfo{person}{Lukas Lerche}.}
  \bibinfo{year}{2017}\natexlab{}.
\newblock \showarticletitle{Efficient optimization of multiple recommendation
  quality factors according to individual user tendencies}.
\newblock \bibinfo{journal}{\emph{Expert Syst. Appl.}}  \bibinfo{volume}{81}
  (\bibinfo{year}{2017}), \bibinfo{pages}{321--331}.
\newblock
\urldef\tempurl%
\url{https://doi.org/10.1016/j.eswa.2017.03.055}
\showDOI{\tempurl}


\bibitem[Kang et~al\mbox{.}(2012)]%
        {10.1145/2124295.2124350}
\bibfield{author}{\bibinfo{person}{Changsung Kang}, \bibinfo{person}{Xuanhui
  Wang}, \bibinfo{person}{Yi Chang}, {and} \bibinfo{person}{Belle Tseng}.}
  \bibinfo{year}{2012}\natexlab{}.
\newblock \showarticletitle{Learning to Rank with Multi-Aspect Relevance for
  Vertical Search}. In \bibinfo{booktitle}{\emph{Proceedings of the Fifth ACM
  International Conference on Web Search and Data Mining}} (Seattle,
  Washington, USA) \emph{(\bibinfo{series}{WSDM '12})}.
  \bibinfo{publisher}{Association for Computing Machinery},
  \bibinfo{address}{New York, NY, USA}, \bibinfo{pages}{453–462}.
\newblock
\showISBNx{9781450307475}
\urldef\tempurl%
\url{https://doi.org/10.1145/2124295.2124350}
\showDOI{\tempurl}


\bibitem[Karmaker~Santu et~al\mbox{.}(2017)]%
        {10.1145/3077136.3080838}
\bibfield{author}{\bibinfo{person}{Shubhra~Kanti Karmaker~Santu},
  \bibinfo{person}{Parikshit Sondhi}, {and} \bibinfo{person}{ChengXiang Zhai}.}
  \bibinfo{year}{2017}\natexlab{}.
\newblock \showarticletitle{On Application of Learning to Rank for E-Commerce
  Search}. In \bibinfo{booktitle}{\emph{Proceedings of the 40th International
  ACM SIGIR Conference on Research and Development in Information Retrieval}}
  (Shinjuku, Tokyo, Japan) \emph{(\bibinfo{series}{SIGIR '17})}.
  \bibinfo{publisher}{Association for Computing Machinery},
  \bibinfo{address}{New York, NY, USA}, \bibinfo{pages}{475–484}.
\newblock
\showISBNx{9781450350228}
\urldef\tempurl%
\url{https://doi.org/10.1145/3077136.3080838}
\showDOI{\tempurl}


\bibitem[Ke et~al\mbox{.}(2017)]%
        {10.5555/3294996.3295074}
\bibfield{author}{\bibinfo{person}{Guolin Ke}, \bibinfo{person}{Qi Meng},
  \bibinfo{person}{Thomas Finley}, \bibinfo{person}{Taifeng Wang},
  \bibinfo{person}{Wei Chen}, \bibinfo{person}{Weidong Ma},
  \bibinfo{person}{Qiwei Ye}, {and} \bibinfo{person}{Tie-Yan Liu}.}
  \bibinfo{year}{2017}\natexlab{}.
\newblock \showarticletitle{LightGBM: A Highly Efficient Gradient Boosting
  Decision Tree}. In \bibinfo{booktitle}{\emph{Proceedings of the 31st
  International Conference on Neural Information Processing Systems}} (Long
  Beach, California, USA) \emph{(\bibinfo{series}{NIPS'17})}.
  \bibinfo{publisher}{Curran Associates Inc.}, \bibinfo{address}{Red Hook, NY,
  USA}, \bibinfo{pages}{3149–3157}.
\newblock
\showISBNx{9781510860964}


\bibitem[Knowles and Corne(2003)]%
        {DBLP:journals/tec/KnowlesC03}
\bibfield{author}{\bibinfo{person}{Joshua~D. Knowles} {and}
  \bibinfo{person}{David Corne}.} \bibinfo{year}{2003}\natexlab{}.
\newblock \showarticletitle{Properties of an adaptive archiving algorithm for
  storing nondominated vectors}.
\newblock \bibinfo{journal}{\emph{{IEEE} Trans. Evol. Comput.}}
  \bibinfo{volume}{7}, \bibinfo{number}{2} (\bibinfo{year}{2003}),
  \bibinfo{pages}{100--116}.
\newblock
\urldef\tempurl%
\url{https://doi.org/10.1109/TEVC.2003.810755}
\showDOI{\tempurl}


\bibitem[Leake(2001)]%
        {DBLP:journals/jors/Leake01}
\bibfield{author}{\bibinfo{person}{Charles~R. Leake}.}
  \bibinfo{year}{2001}\natexlab{}.
\newblock \showarticletitle{Multicriterion Decision in Management: Principles
  and Practice}.
\newblock \bibinfo{journal}{\emph{J. Oper. Res. Soc.}} \bibinfo{volume}{52},
  \bibinfo{number}{5} (\bibinfo{year}{2001}), \bibinfo{pages}{603}.
\newblock
\urldef\tempurl%
\url{https://doi.org/10.1057/palgrave.jors.2601200}
\showDOI{\tempurl}


\bibitem[Li and Yao(2019)]%
        {DBLP:journals/csur/LiY19}
\bibfield{author}{\bibinfo{person}{Miqing Li} {and} \bibinfo{person}{Xin Yao}.}
  \bibinfo{year}{2019}\natexlab{}.
\newblock \showarticletitle{Quality Evaluation of Solution Sets in
  Multiobjective Optimisation: {A} Survey}.
\newblock \bibinfo{journal}{\emph{{ACM} Comput. Surv.}} \bibinfo{volume}{52},
  \bibinfo{number}{2} (\bibinfo{year}{2019}), \bibinfo{pages}{26:1--26:38}.
\newblock
\urldef\tempurl%
\url{https://doi.org/10.1145/3300148}
\showDOI{\tempurl}


\bibitem[Lightner and Director(1981)]%
        {1084969}
\bibfield{author}{\bibinfo{person}{M. Lightner} {and} \bibinfo{person}{S.
  Director}.} \bibinfo{year}{1981}\natexlab{}.
\newblock \showarticletitle{Multiple criterion optimization for the design of
  electronic circuits}.
\newblock \bibinfo{journal}{\emph{IEEE Transactions on Circuits and Systems}}
  \bibinfo{volume}{28}, \bibinfo{number}{3} (\bibinfo{year}{1981}),
  \bibinfo{pages}{169--179}.
\newblock
\urldef\tempurl%
\url{https://doi.org/10.1109/TCS.1981.1084969}
\showDOI{\tempurl}


\bibitem[Lin et~al\mbox{.}(2019)]%
        {DBLP:conf/recsys/LinCPSXSZOJ19}
\bibfield{author}{\bibinfo{person}{Xiao Lin}, \bibinfo{person}{Hongjie Chen},
  \bibinfo{person}{Changhua Pei}, \bibinfo{person}{Fei Sun},
  \bibinfo{person}{Xuanji Xiao}, \bibinfo{person}{Hanxiao Sun},
  \bibinfo{person}{Yongfeng Zhang}, \bibinfo{person}{Wenwu Ou}, {and}
  \bibinfo{person}{Peng Jiang}.} \bibinfo{year}{2019}\natexlab{}.
\newblock \showarticletitle{A pareto-efficient algorithm for multiple objective
  optimization in e-commerce recommendation}. In
  \bibinfo{booktitle}{\emph{Proceedings of the 13th {ACM} Conference on
  Recommender Systems, RecSys 2019, Copenhagen, Denmark, September 16-20,
  2019}}, \bibfield{editor}{\bibinfo{person}{Toine Bogers},
  \bibinfo{person}{Alan Said}, \bibinfo{person}{Peter Brusilovsky}, {and}
  \bibinfo{person}{Domonkos Tikk}} (Eds.). \bibinfo{publisher}{{ACM}},
  \bibinfo{pages}{20--28}.
\newblock
\urldef\tempurl%
\url{https://doi.org/10.1145/3298689.3346998}
\showDOI{\tempurl}


\bibitem[Long et~al\mbox{.}(2012)]%
        {10.1145/2396761.2398671}
\bibfield{author}{\bibinfo{person}{Bo Long}, \bibinfo{person}{Jiang Bian},
  \bibinfo{person}{Anlei Dong}, {and} \bibinfo{person}{Yi Chang}.}
  \bibinfo{year}{2012}\natexlab{}.
\newblock \showarticletitle{Enhancing Product Search by Best-Selling Prediction
  in e-Commerce}. In \bibinfo{booktitle}{\emph{Proceedings of the 21st ACM
  International Conference on Information and Knowledge Management}} (Maui,
  Hawaii, USA) \emph{(\bibinfo{series}{CIKM '12})}.
  \bibinfo{publisher}{Association for Computing Machinery},
  \bibinfo{address}{New York, NY, USA}, \bibinfo{pages}{2479–2482}.
\newblock
\showISBNx{9781450311564}
\urldef\tempurl%
\url{https://doi.org/10.1145/2396761.2398671}
\showDOI{\tempurl}


\bibitem[Lucchese et~al\mbox{.}(2015)]%
        {SIGIR2015}
\bibfield{author}{\bibinfo{person}{Claudio Lucchese},
  \bibinfo{person}{Franco~Maria Nardini}, \bibinfo{person}{Salvatore Orlando},
  \bibinfo{person}{Raffaele Perego}, \bibinfo{person}{Nicola Tonellotto}, {and}
  \bibinfo{person}{Rossano Venturini}.} \bibinfo{year}{2015}\natexlab{}.
\newblock \showarticletitle{QuickScorer: A Fast Algorithm to Rank Documents
  with Additive Ensembles of Regression Trees}. In
  \bibinfo{booktitle}{\emph{Proc. ACM SIGIR}}. \bibinfo{pages}{73--82}.
\newblock


\bibitem[Marler and Arora(2004)]%
        {moosurvey}
\bibfield{author}{\bibinfo{person}{R. Marler} {and} \bibinfo{person}{Jasbir
  Arora}.} \bibinfo{year}{2004}\natexlab{}.
\newblock \showarticletitle{Survey of Multi-Objective Optimization Methods for
  Engineering}.
\newblock \bibinfo{journal}{\emph{Structural and Multidisciplinary
  Optimization}}  \bibinfo{volume}{26} (\bibinfo{date}{04}
  \bibinfo{year}{2004}), \bibinfo{pages}{369--395}.
\newblock
\urldef\tempurl%
\url{https://doi.org/10.1007/s00158-003-0368-6}
\showDOI{\tempurl}


\bibitem[Naghiaei et~al\mbox{.}(2022)]%
        {DBLP:conf/sigir/NaghiaeiRD22}
\bibfield{author}{\bibinfo{person}{Mohammadmehdi Naghiaei},
  \bibinfo{person}{Hossein~A. Rahmani}, {and} \bibinfo{person}{Yashar
  Deldjoo}.} \bibinfo{year}{2022}\natexlab{}.
\newblock \showarticletitle{CPFair: Personalized Consumer and Producer Fairness
  Re-ranking for Recommender Systems}. In \bibinfo{booktitle}{\emph{{SIGIR}
  '22: The 45th International {ACM} {SIGIR} Conference on Research and
  Development in Information Retrieval, Madrid, Spain, July 11 - 15, 2022}},
  \bibfield{editor}{\bibinfo{person}{Enrique Amig{\'{o}}},
  \bibinfo{person}{Pablo Castells}, \bibinfo{person}{Julio Gonzalo},
  \bibinfo{person}{Ben Carterette}, \bibinfo{person}{J.~Shane Culpepper}, {and}
  \bibinfo{person}{Gabriella Kazai}} (Eds.). \bibinfo{publisher}{{ACM}},
  \bibinfo{pages}{770--779}.
\newblock
\urldef\tempurl%
\url{https://doi.org/10.1145/3477495.3531959}
\showDOI{\tempurl}


\bibitem[Nardini et~al\mbox{.}(2022)]%
        {nardini2022distilled}
\bibfield{author}{\bibinfo{person}{Franco~Maria Nardini},
  \bibinfo{person}{Cosimo Rulli}, \bibinfo{person}{Salvatore Trani}, {and}
  \bibinfo{person}{Rossano Venturini}.} \bibinfo{year}{2022}\natexlab{}.
\newblock \showarticletitle{Distilled Neural Networks for Efficient Learning to
  Rank}.
\newblock \bibinfo{journal}{\emph{IEEE Transactions on Knowledge and Data
  Engineering}} (\bibinfo{year}{2022}).
\newblock


\bibitem[Noia et~al\mbox{.}(2017)]%
        {DBLP:journals/isci/NoiaRTS17}
\bibfield{author}{\bibinfo{person}{Tommaso~Di Noia}, \bibinfo{person}{Jessica
  Rosati}, \bibinfo{person}{Paolo Tomeo}, {and} \bibinfo{person}{Eugenio~Di
  Sciascio}.} \bibinfo{year}{2017}\natexlab{}.
\newblock \showarticletitle{Adaptive multi-attribute diversity for recommender
  systems}.
\newblock \bibinfo{journal}{\emph{Inf. Sci.}}  \bibinfo{volume}{382-383}
  (\bibinfo{year}{2017}), \bibinfo{pages}{234--253}.
\newblock
\urldef\tempurl%
\url{https://doi.org/10.1016/j.ins.2016.11.015}
\showDOI{\tempurl}


\bibitem[Oh et~al\mbox{.}(2011)]%
        {DBLP:conf/icdm/OhPYSP11}
\bibfield{author}{\bibinfo{person}{Jinoh Oh}, \bibinfo{person}{Sun Park},
  \bibinfo{person}{Hwanjo Yu}, \bibinfo{person}{Min Song}, {and}
  \bibinfo{person}{Seung{-}Taek Park}.} \bibinfo{year}{2011}\natexlab{}.
\newblock \showarticletitle{Novel Recommendation Based on Personal Popularity
  Tendency}. In \bibinfo{booktitle}{\emph{11th {IEEE} International Conference
  on Data Mining, {ICDM} 2011, Vancouver, BC, Canada, December 11-14, 2011}},
  \bibfield{editor}{\bibinfo{person}{Diane~J. Cook}, \bibinfo{person}{Jian
  Pei}, \bibinfo{person}{Wei Wang}, \bibinfo{person}{Osmar~R. Za{\"{\i}}ane},
  {and} \bibinfo{person}{Xindong Wu}} (Eds.). \bibinfo{publisher}{{IEEE}
  Computer Society}, \bibinfo{pages}{507--516}.
\newblock
\urldef\tempurl%
\url{https://doi.org/10.1109/ICDM.2011.110}
\showDOI{\tempurl}


\bibitem[Paszke et~al\mbox{.}(2019)]%
        {DBLP:conf/nips/PaszkeGMLBCKLGA19}
\bibfield{author}{\bibinfo{person}{Adam Paszke}, \bibinfo{person}{Sam Gross},
  \bibinfo{person}{Francisco Massa}, \bibinfo{person}{Adam Lerer},
  \bibinfo{person}{James Bradbury}, \bibinfo{person}{Gregory Chanan},
  \bibinfo{person}{Trevor Killeen}, \bibinfo{person}{Zeming Lin},
  \bibinfo{person}{Natalia Gimelshein}, \bibinfo{person}{Luca Antiga},
  \bibinfo{person}{Alban Desmaison}, \bibinfo{person}{Andreas K{\"{o}}pf},
  \bibinfo{person}{Edward~Z. Yang}, \bibinfo{person}{Zachary DeVito},
  \bibinfo{person}{Martin Raison}, \bibinfo{person}{Alykhan Tejani},
  \bibinfo{person}{Sasank Chilamkurthy}, \bibinfo{person}{Benoit Steiner},
  \bibinfo{person}{Lu Fang}, \bibinfo{person}{Junjie Bai}, {and}
  \bibinfo{person}{Soumith Chintala}.} \bibinfo{year}{2019}\natexlab{}.
\newblock \showarticletitle{PyTorch: An Imperative Style, High-Performance Deep
  Learning Library}. In \bibinfo{booktitle}{\emph{Advances in Neural
  Information Processing Systems 32: Annual Conference on Neural Information
  Processing Systems 2019, NeurIPS 2019, December 8-14, 2019, Vancouver, BC,
  Canada}}, \bibfield{editor}{\bibinfo{person}{Hanna~M. Wallach},
  \bibinfo{person}{Hugo Larochelle}, \bibinfo{person}{Alina Beygelzimer},
  \bibinfo{person}{Florence d'Alch{\'{e}}{-}Buc}, \bibinfo{person}{Emily~B.
  Fox}, {and} \bibinfo{person}{Roman Garnett}} (Eds.).
  \bibinfo{pages}{8024--8035}.
\newblock
\urldef\tempurl%
\url{https://proceedings.neurips.cc/paper/2019/hash/bdbca288fee7f92f2bfa9f7012727740-Abstract.html}
\showURL{%
\tempurl}


\bibitem[Pobrotyn and Bialobrzeski(2021)]%
        {DBLP:journals/corr/abs-2102-07831}
\bibfield{author}{\bibinfo{person}{Przemyslaw Pobrotyn} {and}
  \bibinfo{person}{Radoslaw Bialobrzeski}.} \bibinfo{year}{2021}\natexlab{}.
\newblock \showarticletitle{NeuralNDCG: Direct Optimisation of a Ranking Metric
  via Differentiable Relaxation of Sorting}.
\newblock \bibinfo{journal}{\emph{CoRR}}  \bibinfo{volume}{abs/2102.07831}
  (\bibinfo{year}{2021}).
\newblock
\showeprint[arXiv]{2102.07831}
\urldef\tempurl%
\url{https://arxiv.org/abs/2102.07831}
\showURL{%
\tempurl}


\bibitem[Qin and Liu(2013)]%
        {DBLP:journals/corr/QinL13}
\bibfield{author}{\bibinfo{person}{Tao Qin} {and} \bibinfo{person}{Tie{-}Yan
  Liu}.} \bibinfo{year}{2013}\natexlab{}.
\newblock \showarticletitle{Introducing {LETOR} 4.0 Datasets}.
\newblock \bibinfo{journal}{\emph{CoRR}}  \bibinfo{volume}{abs/1306.2597}
  (\bibinfo{year}{2013}).
\newblock
\showeprint[arXiv]{1306.2597}
\urldef\tempurl%
\url{http://arxiv.org/abs/1306.2597}
\showURL{%
\tempurl}


\bibitem[Ribeiro et~al\mbox{.}(2012)]%
        {DBLP:conf/recsys/RibeiroLVZ12}
\bibfield{author}{\bibinfo{person}{Marco~T{\'{u}}lio Ribeiro},
  \bibinfo{person}{An{\'{\i}}sio Lacerda}, \bibinfo{person}{Adriano Veloso},
  {and} \bibinfo{person}{Nivio Ziviani}.} \bibinfo{year}{2012}\natexlab{}.
\newblock \showarticletitle{Pareto-efficient hybridization for multi-objective
  recommender systems}. In \bibinfo{booktitle}{\emph{Sixth {ACM} Conference on
  Recommender Systems, RecSys '12, Dublin, Ireland, September 9-13, 2012}},
  \bibfield{editor}{\bibinfo{person}{Padraig Cunningham},
  \bibinfo{person}{Neil~J. Hurley}, \bibinfo{person}{Ido Guy}, {and}
  \bibinfo{person}{Sarabjot~Singh Anand}} (Eds.). \bibinfo{publisher}{{ACM}},
  \bibinfo{pages}{19--26}.
\newblock
\urldef\tempurl%
\url{https://doi.org/10.1145/2365952.2365962}
\showDOI{\tempurl}


\bibitem[Santos et~al\mbox{.}(2015)]%
        {INR-040}
\bibfield{author}{\bibinfo{person}{Rodrygo L.~T. Santos},
  \bibinfo{person}{Craig Macdonald}, {and} \bibinfo{person}{Iadh Ounis}.}
  \bibinfo{year}{2015}\natexlab{}.
\newblock \showarticletitle{Search Result Diversification}.
\newblock \bibinfo{journal}{\emph{Foundations and Trends® in Information
  Retrieval}} \bibinfo{volume}{9}, \bibinfo{number}{1} (\bibinfo{year}{2015}),
  \bibinfo{pages}{1--90}.
\newblock
\showISSN{1554-0669}
\urldef\tempurl%
\url{https://doi.org/10.1561/1500000040}
\showDOI{\tempurl}


\bibitem[Stamenkovic et~al\mbox{.}(2022)]%
        {DBLP:conf/wsdm/StamenkovicKAXK22}
\bibfield{author}{\bibinfo{person}{Dusan Stamenkovic},
  \bibinfo{person}{Alexandros Karatzoglou}, \bibinfo{person}{Ioannis Arapakis},
  \bibinfo{person}{Xin Xin}, {and} \bibinfo{person}{Kleomenis Katevas}.}
  \bibinfo{year}{2022}\natexlab{}.
\newblock \showarticletitle{Choosing the Best of Both Worlds: Diverse and Novel
  Recommendations through Multi-Objective Reinforcement Learning}. In
  \bibinfo{booktitle}{\emph{{WSDM} '22: The Fifteenth {ACM} International
  Conference on Web Search and Data Mining, Virtual Event / Tempe, AZ, USA,
  February 21 - 25, 2022}}, \bibfield{editor}{\bibinfo{person}{K.~Selcuk
  Candan}, \bibinfo{person}{Huan Liu}, \bibinfo{person}{Leman Akoglu},
  \bibinfo{person}{Xin~Luna Dong}, {and} \bibinfo{person}{Jiliang Tang}}
  (Eds.). \bibinfo{publisher}{{ACM}}, \bibinfo{pages}{957--965}.
\newblock
\urldef\tempurl%
\url{https://doi.org/10.1145/3488560.3498471}
\showDOI{\tempurl}


\bibitem[Steck(2018)]%
        {DBLP:conf/recsys/Steck18}
\bibfield{author}{\bibinfo{person}{Harald Steck}.}
  \bibinfo{year}{2018}\natexlab{}.
\newblock \showarticletitle{Calibrated recommendations}. In
  \bibinfo{booktitle}{\emph{Proceedings of the 12th {ACM} Conference on
  Recommender Systems, RecSys 2018, Vancouver, BC, Canada, October 2-7, 2018}},
  \bibfield{editor}{\bibinfo{person}{Sole Pera}, \bibinfo{person}{Michael~D.
  Ekstrand}, \bibinfo{person}{Xavier Amatriain}, {and} \bibinfo{person}{John
  O'Donovan}} (Eds.). \bibinfo{publisher}{{ACM}}, \bibinfo{pages}{154--162}.
\newblock
\urldef\tempurl%
\url{https://doi.org/10.1145/3240323.3240372}
\showDOI{\tempurl}


\bibitem[Steck(2019)]%
        {DBLP:conf/www/Steck19}
\bibfield{author}{\bibinfo{person}{Harald Steck}.}
  \bibinfo{year}{2019}\natexlab{}.
\newblock \showarticletitle{Embarrassingly Shallow Autoencoders for Sparse
  Data}. In \bibinfo{booktitle}{\emph{The World Wide Web Conference, {WWW}
  2019, San Francisco, CA, USA, May 13-17, 2019}},
  \bibfield{editor}{\bibinfo{person}{Ling Liu}, \bibinfo{person}{Ryen~W.
  White}, \bibinfo{person}{Amin Mantrach}, \bibinfo{person}{Fabrizio
  Silvestri}, \bibinfo{person}{Julian~J. McAuley}, \bibinfo{person}{Ricardo
  Baeza{-}Yates}, {and} \bibinfo{person}{Leila Zia}} (Eds.).
  \bibinfo{publisher}{{ACM}}, \bibinfo{pages}{3251--3257}.
\newblock
\urldef\tempurl%
\url{https://doi.org/10.1145/3308558.3313710}
\showDOI{\tempurl}


\bibitem[Svore et~al\mbox{.}(2011)]%
        {10.1145/1963405.1963459}
\bibfield{author}{\bibinfo{person}{Krysta~M. Svore},
  \bibinfo{person}{Maksims~N. Volkovs}, {and} \bibinfo{person}{Christopher~J.C.
  Burges}.} \bibinfo{year}{2011}\natexlab{}.
\newblock \showarticletitle{Learning to Rank with Multiple Objective
  Functions}. In \bibinfo{booktitle}{\emph{Proceedings of the 20th
  International Conference on World Wide Web}} (Hyderabad, India)
  \emph{(\bibinfo{series}{WWW '11})}. \bibinfo{publisher}{Association for
  Computing Machinery}, \bibinfo{address}{New York, NY, USA},
  \bibinfo{pages}{367–376}.
\newblock
\showISBNx{9781450306324}
\urldef\tempurl%
\url{https://doi.org/10.1145/1963405.1963459}
\showDOI{\tempurl}


\bibitem[van Doorn et~al\mbox{.}(2016)]%
        {10.1145/2911451.2914708}
\bibfield{author}{\bibinfo{person}{Joost van Doorn}, \bibinfo{person}{Daan
  Odijk}, \bibinfo{person}{Diederik~M. Roijers}, {and} \bibinfo{person}{Maarten
  de Rijke}.} \bibinfo{year}{2016}\natexlab{}.
\newblock \showarticletitle{Balancing Relevance Criteria through
  Multi-Objective Optimization}. In \bibinfo{booktitle}{\emph{Proceedings of
  the 39th International ACM SIGIR Conference on Research and Development in
  Information Retrieval}} (Pisa, Italy) \emph{(\bibinfo{series}{SIGIR '16})}.
  \bibinfo{publisher}{Association for Computing Machinery},
  \bibinfo{address}{New York, NY, USA}, \bibinfo{pages}{769–772}.
\newblock
\showISBNx{9781450340694}
\urldef\tempurl%
\url{https://doi.org/10.1145/2911451.2914708}
\showDOI{\tempurl}


\bibitem[Vargas and Castells(2011)]%
        {DBLP:conf/recsys/VargasC11}
\bibfield{author}{\bibinfo{person}{Saul Vargas} {and} \bibinfo{person}{Pablo
  Castells}.} \bibinfo{year}{2011}\natexlab{}.
\newblock \showarticletitle{Rank and relevance in novelty and diversity metrics
  for recommender systems}. In \bibinfo{booktitle}{\emph{Proceedings of the
  2011 {ACM} Conference on Recommender Systems, RecSys 2011, Chicago, IL, USA,
  October 23-27, 2011}}, \bibfield{editor}{\bibinfo{person}{Bamshad Mobasher},
  \bibinfo{person}{Robin~D. Burke}, \bibinfo{person}{Dietmar Jannach}, {and}
  \bibinfo{person}{Gediminas Adomavicius}} (Eds.). \bibinfo{publisher}{{ACM}},
  \bibinfo{pages}{109--116}.
\newblock
\urldef\tempurl%
\url{https://dl.acm.org/citation.cfm?id=2043955}
\showURL{%
\tempurl}


\bibitem[von L{\"{u}}cken et~al\mbox{.}(2014)]%
        {DBLP:journals/coap/LuckenBB14}
\bibfield{author}{\bibinfo{person}{Christian von L{\"{u}}cken},
  \bibinfo{person}{Benjam{\'{\i}}n Bar{\'{a}}n}, {and}
  \bibinfo{person}{Carlos~A. Brizuela}.} \bibinfo{year}{2014}\natexlab{}.
\newblock \showarticletitle{A survey on multi-objective evolutionary algorithms
  for many-objective problems}.
\newblock \bibinfo{journal}{\emph{Comput. Optim. Appl.}} \bibinfo{volume}{58},
  \bibinfo{number}{3} (\bibinfo{year}{2014}), \bibinfo{pages}{707--756}.
\newblock
\urldef\tempurl%
\url{https://doi.org/10.1007/s10589-014-9644-1}
\showDOI{\tempurl}


\bibitem[Wan et~al\mbox{.}(2019)]%
        {DBLP:conf/acl/WanMNM19}
\bibfield{author}{\bibinfo{person}{Mengting Wan}, \bibinfo{person}{Rishabh
  Misra}, \bibinfo{person}{Ndapa Nakashole}, {and} \bibinfo{person}{Julian~J.
  McAuley}.} \bibinfo{year}{2019}\natexlab{}.
\newblock \showarticletitle{Fine-Grained Spoiler Detection from Large-Scale
  Review Corpora}. In \bibinfo{booktitle}{\emph{Proceedings of the 57th
  Conference of the Association for Computational Linguistics, {ACL} 2019,
  Florence, Italy, July 28- August 2, 2019, Volume 1: Long Papers}},
  \bibfield{editor}{\bibinfo{person}{Anna Korhonen}, \bibinfo{person}{David~R.
  Traum}, {and} \bibinfo{person}{Llu{\'{\i}}s M{\`{a}}rquez}} (Eds.).
  \bibinfo{publisher}{Association for Computational Linguistics},
  \bibinfo{pages}{2605--2610}.
\newblock
\urldef\tempurl%
\url{https://doi.org/10.18653/v1/p19-1248}
\showDOI{\tempurl}


\bibitem[Wang et~al\mbox{.}(2012)]%
        {10.1145/2348283.2348385}
\bibfield{author}{\bibinfo{person}{Lidan Wang}, \bibinfo{person}{Paul~N.
  Bennett}, {and} \bibinfo{person}{Kevyn Collins-Thompson}.}
  \bibinfo{year}{2012}\natexlab{}.
\newblock \showarticletitle{Robust Ranking Models via Risk-Sensitive
  Optimization}. In \bibinfo{booktitle}{\emph{Proceedings of the 35th
  International ACM SIGIR Conference on Research and Development in Information
  Retrieval}} (Portland, Oregon, USA) \emph{(\bibinfo{series}{SIGIR '12})}.
  \bibinfo{publisher}{Association for Computing Machinery},
  \bibinfo{address}{New York, NY, USA}, \bibinfo{pages}{761–770}.
\newblock
\showISBNx{9781450314725}
\urldef\tempurl%
\url{https://doi.org/10.1145/2348283.2348385}
\showDOI{\tempurl}


\bibitem[Wang et~al\mbox{.}(2016)]%
        {DBLP:journals/kbs/WangGLY16}
\bibfield{author}{\bibinfo{person}{Shanfeng Wang}, \bibinfo{person}{Maoguo
  Gong}, \bibinfo{person}{Haoliang Li}, {and} \bibinfo{person}{Junwei Yang}.}
  \bibinfo{year}{2016}\natexlab{}.
\newblock \showarticletitle{Multi-objective optimization for long tail
  recommendation}.
\newblock \bibinfo{journal}{\emph{Knowl. Based Syst.}}  \bibinfo{volume}{104}
  (\bibinfo{year}{2016}), \bibinfo{pages}{145--155}.
\newblock
\urldef\tempurl%
\url{https://doi.org/10.1016/j.knosys.2016.04.018}
\showDOI{\tempurl}


\bibitem[Wu et~al\mbox{.}(2022)]%
        {wu2022multifr}
\bibfield{author}{\bibinfo{person}{Haolun Wu}, \bibinfo{person}{Chen Ma},
  \bibinfo{person}{Bhaskar Mitra}, \bibinfo{person}{Fernando Diaz}, {and}
  \bibinfo{person}{Xue Liu}.} \bibinfo{year}{2022}\natexlab{}.
\newblock \showarticletitle{Multi-FR: A Multi-objective Optimization Framework
  for Multi-stakeholder Fairness-aware Recommendation}. In
  \bibinfo{booktitle}{\emph{Transactions on Information Systems (TOIS)}}.
  \bibinfo{publisher}{{ACM}}.
\newblock


\bibitem[Wu et~al\mbox{.}(2010)]%
        {lambdamart}
\bibfield{author}{\bibinfo{person}{Q. Wu}, \bibinfo{person}{C.J.C. Burges},
  \bibinfo{person}{K.M. Svore}, {and} \bibinfo{person}{J. Gao}.}
  \bibinfo{year}{2010}\natexlab{}.
\newblock \showarticletitle{Adapting boosting for information retrieval
  measures}.
\newblock \bibinfo{journal}{\emph{Information Retrieval}}
  (\bibinfo{year}{2010}).
\newblock


\bibitem[Zheng and Wang(2022)]%
        {DBLP:journals/ijon/ZhengW22}
\bibfield{author}{\bibinfo{person}{Yong Zheng} {and}
  \bibinfo{person}{David~(Xuejun) Wang}.} \bibinfo{year}{2022}\natexlab{}.
\newblock \showarticletitle{A survey of recommender systems with
  multi-objective optimization}.
\newblock \bibinfo{journal}{\emph{Neurocomputing}}  \bibinfo{volume}{474}
  (\bibinfo{year}{2022}), \bibinfo{pages}{141--153}.
\newblock
\urldef\tempurl%
\url{https://doi.org/10.1016/j.neucom.2021.11.041}
\showDOI{\tempurl}


\bibitem[Zitzler et~al\mbox{.}(2007)]%
        {DBLP:conf/emo/ZitzlerBT06}
\bibfield{author}{\bibinfo{person}{Eckart Zitzler}, \bibinfo{person}{Dimo
  Brockhoff}, {and} \bibinfo{person}{Lothar Thiele}.}
  \bibinfo{year}{2007}\natexlab{}.
\newblock \showarticletitle{The Hypervolume Indicator Revisited: On the Design
  of Pareto-compliant Indicators Via Weighted Integration}. In
  \bibinfo{booktitle}{\emph{Evolutionary Multi-Criterion Optimization, 4th
  International Conference, {EMO} 2007, Matsushima, Japan, March 5-8, 2007,
  Proceedings}} \emph{(\bibinfo{series}{Lecture Notes in Computer Science},
  Vol.~\bibinfo{volume}{4403})}, \bibfield{editor}{\bibinfo{person}{Shigeru
  Obayashi}, \bibinfo{person}{Kalyanmoy Deb}, \bibinfo{person}{Carlo Poloni},
  \bibinfo{person}{Tomoyuki Hiroyasu}, {and} \bibinfo{person}{Tadahiko Murata}}
  (Eds.). \bibinfo{publisher}{Springer}, \bibinfo{pages}{862--876}.
\newblock
\urldef\tempurl%
\url{https://doi.org/10.1007/978-3-540-70928-2\_64}
\showDOI{\tempurl}


\end{thebibliography}

\end{document}